\documentclass[prd,preprint,superscriptaddress,
  tightenlines,nofootinbib, eqsecnum
  ]{revtex4}

\usepackage{amsmath}
\usepackage{yfonts}
\usepackage[usenames,dvipsnames]{xcolor}
\usepackage{ytableau}
\usepackage{graphicx}
\usepackage{tensor}
\usepackage[utf8]{inputenc}

\newcommand{\dd}{\mathrm{d}}
\newcommand{\di}{\mathrm{i}}

\newcommand{\It}{\widetilde{I}}
\newcommand{\Jt}{\widetilde{J}}
\newcommand{\Kt}{\widetilde{K}}
\newcommand{\Wt}{\widetilde{W}}
\newcommand{\Xt}{\widetilde{X}}
\newcommand{\Yt}{\widetilde{Y}}
\newcommand{\Zt}{\widetilde{Z}}
\newcommand{\Pt}{\widetilde{P}}
\newcommand{\Qzt}{\widetilde{Q}^{(0)}}
\newcommand{\Qmt}{\widetilde{Q}^{(-)}}
\newcommand{\Qpt}{\widetilde{Q}^{(+)}}
\newcommand{\dQmt}{\dot{\widetilde{Q}}^{(-)}}
\newcommand{\ddQmt}{\ddot{\widetilde{Q}}^{(-)}}

\usepackage{amsmath}
\usepackage{amsfonts}
\usepackage{amssymb}
\usepackage{bm}
\usepackage{hyperref}
\usepackage{mathrsfs}
\usepackage{graphicx}

\usepackage{empheq}

\usepackage{ulem}
\definecolor{darkgreen}{rgb}{0,0.5,0}

\hypersetup{
 bookmarks=true,         
 unicode=false,          
 pdftoolbar=true,        
 pdfmenubar=true,        
 pdffitwindow=false,     
 pdfstartview={FitH},    
 pdftitle={My title},    
 pdfauthor={Author},     
 pdfsubject={Subject},   
 pdfcreator={Creator},   
 pdfproducer={Producer}, 
 pdfkeywords={keyword1} {key2} {key3}, 
 pdfnewwindow=true,      
 colorlinks=true,       
 linkcolor=red,          
 citecolor=cyan,        
 filecolor=magenta,      
 urlcolor=darkgreen,           
 linktocpage=true
}

\allowdisplaybreaks

\DeclareSymbolFontAlphabet{\mathrsfs}{rsfs}
\DeclareMathAlphabet{\mathcal}{OMS}{cmsy}{m}{n}

\newcommand{\beq}{\begin{equation}}
\newcommand{\eeq}{\end{equation}}

\begin{document}

\title{The current-type quadrupole moment and gravitational-wave mode \boldmath$(\ell,m)=(2,1)$ \unboldmath of compact binary systems \\at the third post-Newtonian order}

\author{Quentin \textsc{Henry}}\email{henry@iap.fr}
\affiliation{$\mathcal{G}\mathbb{R}\varepsilon{\mathbb{C}}\mathcal{O}$, 
Institut d'Astrophysique de Paris,\\ UMR 7095, CNRS, Sorbonne Universit{\'e},\\
98\textsuperscript{bis} boulevard Arago, 75014 Paris, France}

\author{Guillaume \textsc{Faye}}\email{faye@iap.fr}
\affiliation{$\mathcal{G}\mathbb{R}\varepsilon{\mathbb{C}}\mathcal{O}$, 
Institut d'Astrophysique de Paris,\\ UMR 7095, CNRS, Sorbonne Universit{\'e},\\
98\textsuperscript{bis} boulevard Arago, 75014 Paris, France}

\author{Luc \textsc{Blanchet}}\email{luc.blanchet@iap.fr}
\affiliation{$\mathcal{G}\mathbb{R}\varepsilon{\mathbb{C}}\mathcal{O}$, 
Institut d'Astrophysique de Paris,\\ UMR 7095, CNRS, Sorbonne Universit{\'e},\\
98\textsuperscript{bis} boulevard Arago, 75014 Paris, France}

\date{\today}

\begin{abstract} 
Up to the third post-Newtonian (3PN) order, we compute (i) the current-type quadrupole moment of (non-spinning) compact binary systems, as well as (ii) the corresponding gravitational-wave mode $(2,1)$ (constituting a 3.5PN correction in the waveform). Moreover, at this occasion, (iii) we recompute and confirm the previous calculation of the mass-type octupole moment to 3PN order. The ultra-violet (UV) divergences due to the point-like nature of the source are treated by means of dimensional regularization. This entails generalizing the definition of the irreducible mass and current multipole moments of an isolated PN source in $d$ spatial dimensions. In particular, we find that the current type moment has the symmetry of a particular mixed Young tableau and that, in addition, there appears a third type of moment which is however inexistent in 3 spatial dimensions.
\end{abstract}

\pacs{04.25.Nx, 04.30.-w, 97.60.Jd, 97.60.Lf}

\maketitle

\section{Introduction} 
\label{sec:intro}

The power of the gravitational-wave (GW) science is predicated on the precise reconstruction of the GW signal as a function of the parameters of the source. In particular, it is crucial to improve the post-Newtonian (PN) predictions for the amplitude and phase of gravitational waveforms generated by inspiralling compact binaries (see the recent reviews~\cite{BuonSathya15, barack2019black, DHS20}). One of the main challenges for this purpose is the computation of the multipole moments of the compact binary system to high PN order.

Previous computations of multipole moments for compact binaries (without spins) included the mass-type quadrupole moment at 1PN order~\cite{BS89}, 2PN order~\cite{BDI95, WW96, LMRY19}, 2.5PN  order~\cite{B96} and 3PN order~\cite{BIJ02, BI04mult, BDEI05dr}, yielding the fundamental GW mode (2,2) up to 3.5PN order~\cite{K07, BFIS08, FMBI12}. Recently, the 4PN quadrupole moment has been partly obtained in~\cite{MHLMFB20}, but some contributions, coming from the proper treatment of infra-red (IR) divergences by dimensional regularisation, are still missing. The current quadrupole moment has been determined at 1PN order~\cite{DI91a, BDI95, DIN09} and 2.5PN order~\cite{BFIS08}. The mass octupole moment was computed to 3PN order~\cite{FBI15} and the current octupole moment to 2PN order~\cite{DI94, BFIS08}. Other moments are known, such as the mass hexadecapole ($2^4$) one at 2PN order and the $2^5$ mass and $2^4$ current ones at 1PN order~\cite{BFIS08}. Of course, all moments are known in closed form to Newtonian order. High-order PN contributions due to the spins of the compact bodies have also been derived~\cite{BMB13, BFMP15, M15}.

In the present paper, we shall improve the state of affairs by computing the current-type quadrupole moment up to the 3PN order, notably using dimensional regularisation for treating the ultra-violet (UV) divergences associated with the point-like nature of the source (a system of two compact objects modelled as point masses). Therefore, we are led to generalize the wave-generation formalism from isolated PN matter sources as well as the notion of symmetric-trace-free (STF) mass and current multipole moments to arbitrary space dimensions.

In 3 dimensions, the STF property plays a crucial role in this formalism, since it guaranties that the multipole moments are linearly independent. This is the case because they are irreducible tensors, in the sense that they belong to irreducible representations of the rotation group SO(3). Now, the irreducible decomposition of vector and tensor fields in a generic space of dimension $d$ is more complicated than in 3 dimensions. Indeed, irreducible tensors in $d$ dimensions are exactly the trace-free tensors that have Young-tableau symmetries (elements of SO($d$) representations). When $d=3$, because of identities ensuing from the fact that antisymmetrization over more than three indices yields zero (syzygies), any Young tableau can be written as the tensor product of STF and anti-symmetric Levi-Civita tensors. However, for generic $d$, due to the absence of syzygies and the impossibility of generalizing the Levi-Civita tensor, all Young tableaux must a priori be considered.

Regarding the definition of multipole moments in $d$ dimensions, we shall essentially find that, while the mass-type moments admit a straightforward generalization, the current-type moments acquire a non-trivial symmetry described by a mixed Young tableau~\cite{james_1984, ma2004problems, Bekaert:2006py}. In addition, there appears a third type of irreducible moment, absent in 3 dimensions and described by another mixed tableau. Thus, one of the main problem addressed in this paper is that of the irreducible multipole decomposition of the tensorial field of GR in $d$ dimensions. Our work should be useful for the computation of the next-to-leading (5PN) tail effect in the conservative binary's dynamics, which requires the control of the current quadrupole moment in $d$ dimensions~\cite{Foffa:2019eeb, Blumlein21}.

Once the current quadrupole moment is properly defined, we shall apply our standard techniques (described for instance in the exhaustive work~\cite{MHLMFB20}) to compute it with 3PN accuracy in the case of compact binaries without spins. We present the result in the center-of-mass (CM) frame and for quasi-circular orbits. Next, adding corrections coming from non-linear multipole interactions --- most notably the tails-of-tails already known from~\cite{FBI15} ---, we obtain the radiative current quadrupole moment measured at future null infinity. The full physical content of this moment is encoded into the gravitational mode $(\ell,m)=(2,1)$, which we thus provide with 3PN relative accuracy, corresponding to 3.5PN accuracy in the full waveform. The perturbative limit or small mass-ratio limit is an important test for the mode, and found to agree with the result from black-hole perturbation theory~\cite{Sasa94, TSasa94, TTS96}. Our expression for this current quadrupolar mode is ready for comparison with existing numerical relativity calculations such as~\cite{varma2014gravitational, bustillo2015comparison}.

The plan of this paper goes as follows. In Sec.~\ref{sec:currentmult}, we obtain the multipole expansion of the metric generated by an isolated system in generic $d$ dimensions. The final results are given by Eq.~\eqref{ILfinal} for the mass-type moments $I_L$, and by Eqs.~\eqref{JiLfinal} and~\eqref{KijLfinal} for the appropriate generalizations $J_{i\vert L}$ and $K_{ij\vert L}$ of the current-type moments. In Sec.~\ref{sec:currentquad3PN}, we specialize the results to the computation of the current-type quadrupole moment $J_{ij}$ in the case of compact binary systems, to 3PN order. We provide some details on the computation of the most difficult term but, otherwise, refer to~\cite{MHLMFB20} for more precise explanations. Sec.~\ref{sec:mode21} is devoted to the computation of the radiative current quadrupole moment $V_{ij}$ defined at future null infinity, and the corresponding GW mode $h^{21}$, given for quasi-circular orbits by Eq.~\eqref{H21final}. Finally, in Sec.~\eqref{sec:shift}, we perform a partial test of our result by checking the transformation law of the quadrupole moment $J_{ij}$ under a constant shift of the spatial origin, using a formula derived for linearized gravity in Ref.~\cite{DI91a}. Appendix \ref{sec:Young} is a compendium of formulas for the multipole decomposition in $d$ dimensions.

\section{The mass and current multipole moments in \boldmath $d$ \unboldmath dimensions} 
\label{sec:currentmult}

\subsection{Multipole decomposition of the metric of an isolated matter system}

The $d$-dimensional Einstein field equations, relaxed by the harmonic-coordinates condition, take the form of an ordinary wave equation (with $\Box \equiv \eta^{\mu\nu}\partial_\mu\partial_\nu$ the $d$-dimensional flat d'Alembertian operator) for the gothic metric deviation $h^{\mu\nu} \equiv \sqrt{-g}\, g^{\mu\nu}-\eta^{\mu\nu}$:
\begin{equation}\label{EE}
\Box h^{\mu\nu}=\frac{16\pi G}{c^4}\,\tau^{\mu\nu}\,,
\end{equation}
supplemented with the harmonic coordinate condition itself: $\partial_\nu h^{\mu\nu} = 0$. The source of the wave equation is made of the stress-energy pseudo-tensor of matter and gravitational fields
\begin{align}\label{tau}
\tau^{\mu\nu} = \vert g\vert T^{\mu\nu}+\frac{c^4}{16\pi G}\Lambda^{\mu\nu}\,,
\end{align}
where $T^{\mu\nu}$ is the matter stress-energy tensor with spatially compact support (assuming an isolated matter system), and the gravitational stress-energy distribution $\Lambda^{\mu\nu}$ is at least quadratic in $h^{\mu\nu}$ or its first/second space-time derivatives. In this formulation, the matter equations of motion, $\partial_\nu \tau^{\mu\nu} = 0$, are just the consequence of the harmonic coordinate condition.

The retarded solution of Eq.~\eqref{EE} valid in all space-time for a smooth matter distribution and satisfying appropriate (no-incoming wave) boundary conditions at infinity, is given in terms of the $d$-dimensional retarded integral operator $\Box^{-1}_\mathrm{ret}$ as
\begin{align}\label{retsol}
h^{\mu\nu}(\mathbf{x},t) &= \frac{16\pi G}{c^4}\,
\Box^{-1}_\mathrm{ret} \tau^{\mu\nu} \nonumber\\&= - \frac{4 G\,\tilde{k}}{c^4}
\int_1^{+\infty} \dd z \,\gamma_{\frac{1-d}{2}}(z) \int
\dd^d\mathbf{x}'
\,\frac{\tau^{\mu\nu}(\mathbf{x}',t-z\vert\mathbf{x}-\mathbf{x}'\vert/c)}
{\vert\mathbf{x}-\mathbf{x}'\vert^{d-2}}\,.
\end{align}
For convenience, we work in ordinary real space, where the Green function reads, following the notation in~\cite{BBBFMa}:
\begin{align}\label{retgreen} 
G_\text{ret}(\mathbf{x},t) = -
\frac{\tilde{k}}{4\pi}\frac{\theta(c t-r)}{r^{d-1}}
\,\gamma_{\frac{1-d}{2}}\left(\frac{c t}{r}\right)\,,
\end{align}
with $\tilde{k}=\Gamma(\frac{d}{2}-1)/\pi^{\frac{d}{2}-1}$ (so that $\lim_{d\to 3}\tilde{k}=1$) and with the useful definition:
\begin{align}\label{gamma}
&\gamma_{\frac{1-d}{2}}(z) \equiv \frac{2\sqrt{\pi}}{\Gamma(\frac{3-d}{2})\Gamma(\frac{d}{2}-1)}
\,\big(z^2-1\bigr)^{\frac{1-d}{2}}\,,
\end{align}
which is such that $\int_1^{+\infty} \dd z \,\gamma_{\frac{1-d}{2}}(z) = 1$ and $\lim_{d\to 3}\gamma_{\frac{1-d}{2}}(z)=\delta(z-1)$.

Next, we consider the multipole expansion of $h^{\mu\nu}$, denoted as $\mathcal{M}(h^{\mu\nu})$, outside the compact domain of the source. Its general expression has been obtained in 3 dimensions in~\cite{B95, B98mult} and in $d$ dimensions in~\cite{BDEI05dr}. Let us outline the derivation for completeness.  $\mathcal{M}(h^{\mu\nu})$ is a formal solution of the \textit{vacuum} Einstein field equation, \textit{i.e.}, $\Box \mathcal{M}(h^{\mu\nu}) = \mathcal{M}(\Lambda^{\mu\nu})$, where the multipole expansion of the source term is $\mathcal{M}(\Lambda^{\mu\nu})\equiv\Lambda^{\mu\nu}[\mathcal{M}(h^{\mu\nu})]$. This solution obviously agrees with the actual solution $h^{\mu\nu}$ outside the source, and furthermore can be extended inside the source for any $r=\vert\mathbf{x}\vert$, except at $r=0$, where it diverges. To proceed, it is convenient to pose
\begin{equation}\label{defDelta}
h^{\mu\nu} = \mathop{\mathrm{FP}}_{B=0}\,
\Box^{-1}_\mathrm{ret} \Bigl[ \tilde{r}^B \mathcal{M}(\Lambda^{\mu\nu})\Bigr] + \Delta^{\mu\nu} \,.
\end{equation}
The first term is already in the form of a multipole expansion; it is well-defined thanks to a regulator factor $r^B$ and a process of finite part (FP) when $B\to 0$ to cope with the divergence of the multipole expansion at $r=0$. We denote $\tilde{r}=r/r_0$, $r_0$ being some arbitrary constant length scale. Since $\mathcal{M}(h^{\mu\nu})$ solves the vacuum field equations, it follows from~\eqref{defDelta} that $\mathcal{M}(\Delta^{\mu\nu})$ is an homogeneous (retarded) multipolar solution of the wave equation (in $d$ dimensions): $\Box \mathcal{M}(\Delta^{\mu\nu}) = 0$, hence it takes the form
\begin{equation}\label{Delta3}
\mathcal{M}(\Delta^{\mu\nu}) = - \frac{4 G}{c^4}
\sum_{\ell=0}^{+\infty}
\frac{(-)^\ell}{\ell!}\,\partial_L\widetilde{\mathcal{F}}_L^{\mu\nu} \,,
\end{equation}
where $\partial_L\equiv\partial_{i_1}\cdots\partial_{i_\ell}$ is a multi-spatial derivative with respect to the multi-index $L=i_1\cdots i_\ell$ involving $\ell$ indices, and the multipole-moment functions are of the type
\begin{align}\label{tilde}
\widetilde{\mathcal{F}}_L^{\mu\nu}(r,t) =
\frac{\tilde{k}}{r^{d-2}}\int_1^{+\infty} \dd
z\,\gamma_{\frac{1-d}{2}}(z)\,\mathcal{F}_L^{\mu\nu}(t-z r/c)\,,
\end{align}
thus satisfying $\Box\widetilde{\mathcal{F}}_L^{\mu\nu}(r,t)=0$ in $d$ dimensions. In our approach, it is important to impose that $\widetilde{\mathcal{F}}_L^{\mu\nu}$ be symmetric-trace-free (STF) with respect to the $\ell$ indices composing $L$; however the object $\widetilde{\mathcal{F}}_L^{\mu\nu}$ does not represent a genuine irreducible multipole moment because of the additional dependence on the space-time indices $\mu\nu$. The function of time $\mathcal{F}_L^{\mu\nu}(t)$ in~\eqref{tilde} is related to the PN expansion of the pseudo-tensor $\overline{\tau}^{\mu\nu}$ by~\cite{B95, B98mult, BDEI05dr}\footnote{When necessary, we indicate the formal PN expansion of a quantity with an overbar.}
\begin{align}\label{calF0}
\mathcal{F}_L^{\mu\nu}(t)=\mathop{\mathrm{FP}}_{B=0}\int
\dd^d\mathbf{x}\,\tilde{r}^B\,
\hat{x}_L\,\overline{\tau}_{[\ell]}^{\mu\nu}(\mathbf{x},t)\,,
\end{align}
with the STF product of spatial vectors in $d$ dimensions being denoted by $\hat{x}_L\equiv\text{STF}(x_{i_1}\cdots x_{i_\ell})$ or, alternatively, $x_{\langle L\rangle}\equiv\hat{x}_L$ as used below. In turn, the $\ell$-dependent integrand in~\eqref{calF0} can be written down either in ``exact'' form using an intermediate kernel function $\delta_\ell^{(d)} (z)$:
\begin{subequations}\label{exactform}
\begin{align}
\overline{\tau}_{[\ell]}^{\mu\nu}(\mathbf{x},t) &= \int_{-1}^1 \dd z
\,\delta_\ell^{(d)} (z)
\,\overline{\tau}^{\mu\nu}(\mathbf{x},t+z r/c)\,,\\
\text{with}\quad
\delta_\ell^{[d]} (z) &\equiv
\frac{\Gamma\left(\frac{d}{2}+\ell\right)}{
	\sqrt{\pi}\Gamma
	\left(\frac{d-1}{2}+\ell\right)}
\,(1-z^2)^{\frac{d-3}{2}+\ell}\,, \qquad\int_{-1}^{1}
\dd z\,\delta_\ell^{[d]}(z) = 1\,,
\end{align}\end{subequations}
or in the following form, very useful in practice, of the infinite PN series
\begin{subequations}\label{PNform}
\begin{align}
\overline{\tau}_{[\ell]}^{\mu\nu}(\mathbf{x},t) &= \sum_{k=0}^{+\infty}\alpha_\ell^k \left(\frac{r}{c}\frac{\partial}{\partial t}\right)^{2k}\overline{\tau}^{\mu\nu}(\mathbf{x},t)\,,\\ \text{with}\quad\alpha_\ell^k &= \frac{1}{2^{2k}k!}\frac{\Gamma\left(\frac{d}{2}+\ell\right)}{\Gamma\left(\frac{d}{2}+\ell+k\right)}\,.
\end{align}\end{subequations}

\subsection{The irreducible decomposition of the multipole expansion}

Up to this stage, the above derivation is already known. It was actually used to define the mass and current irreducible STF multipole moments $I_L$ and $J_L$ in 3 dimensions, in linearized gravity~\cite{DI91b} and in full GR~\cite{B98mult}. We now come to the crucial point: performing the \textit{irreducible} decomposition of $\mathcal{F}_L^{\mu\nu}$ in $d$ dimensions. 

A previous investigation was already made in~\cite{BDEI05dr}, where the easier case of the mass-type moments $I_L$ was treated. In $d$ dimensions, the mass moments are just STF moments, with the symmetries of symmetric Young tableaux. Now, we shall prove that the appropriate generalization in $d$ dimensions of the current-type moments $J_L$ consists of two and only two additional moments, denoted $J_{i\vert L}$ and $K_{ij\vert L}$, which are also trace-free but have the symmetries of more complicated mixed Young tableaux. Strictly speaking, it is $J_{i\vert L}$ that constitutes the genuine generalization of the usual current moment while $K_{ij\vert L}$ does not exist in 3 dimensions.

To start with, we construct the irreducible decomposition of the scalar, vector and tensor components $00$, $0i$ and $ij$ of the object $\mathcal{F}_L^{\mu\nu}$. The ensuing formulas will generalize the 3-dimensional Eqs.~(5.1)-(5.2) in~\cite{B98mult} and the $d$-dimensional ones~(3.32)-(3.33) in~\cite{BDEI05dr}. We display the results without proofs. For the 00 component, we have nothing to do (since $\mathcal{F}_L^{00}$ is STF in $L$), but, for sake of uniformity, we introduce the specific notation  
\begin{align}\label{decompF00}
\mathcal{F}_L^{00} = R_L \,.
\end{align}
For the $0i$ components, we get, like in 3 dimensions, three irreducible pieces corresponding to STF, ``anti-symmetric'' and trace parts, say
\begin{align}\label{decompF0i}
\mathcal{F}_L^{0i} = T^{(+)}_{iL} + T^{(0)}_{i|\langle i_\ell L-1 \rangle} + \delta_{i \langle i_\ell}
T^{(-)}_{L-1 \rangle} \,.
\end{align}
Here and below, the angular brakets $\langle\cdots\rangle$ surrounding indices represent the STF projection, also indicated by the explicit mention $\text{STF}_L$ (or $\text{STF}_{ij}$), \emph{e.g.}, $T_{i|\langle i_\ell L-1 \rangle} \equiv \text{STF}_L \,T_{i|L}$, with $L=i_1\cdots i_\ell$ and $L-1=i_1\cdots i_{\ell-1}$. The three irreducible pieces in~\eqref{decompF0i} are given by
\begin{subequations}\label{invdecompF0i}
\begin{align}
T^{(+)}_{Li_{\ell+1}} &= \mathcal{F}^{0\langle i_{\ell+1}}_{\phantom{0\langle
		i_{\ell+1}}L \rangle} \,, \\ 
T^{(0)}_{i | i_\ell L-1} &= \frac{2\ell}{\ell+1}
\mathcal{H}^{[i}_{\phantom{[i}i_\ell]L-1} \,, \label{invdecompF0ib}\\
T^{(-)}_{L-1} &= \frac{\ell (2\ell+d-4)}{(\ell+d-3)(2\ell+d-2)} 
\mathcal{F}^{0a}_{aL-1} \,.
\end{align}\end{subequations}
The square brackets $[\cdots]$ mean the anti-symmetrization of the enclosed indices (with the factor $\frac{1}{2}$ included in the present case). As in 3 dimensions, the tensors $T^{(+)}_{Li_{\ell+1}}$ and $T^{(-)}_{L-1}$ are STF in all their indices. However, very importantly, the second piece $T^{(0)}_{i | i_\ell L-1}$ is not, as emphasised by the vertical separation bar. It is defined from the trace-free (TF) projection of $\mathcal{F}_L^{0i}$ over all its indices, \textit{i.e.}, the object $\mathcal{H}^{i}_{L} \equiv \underset{iL}{\text{TF}} \,\mathcal{F}^{0i}_{L}$ (but \textit{without} symmetrization), which reads explicitly
\begin{align}\label{TF1}
\mathcal{H}^i_L &= \mathcal{F}^{0i}_L - \frac{\ell(2\ell+d -4)}{(\ell+d-3)(2\ell+d-2)} \delta_{i\langle i_\ell} \mathcal{F}^{0a}_{L-1\rangle \, a}\,.
\end{align}

The irreducible decomposition of the $ij$ components is still more complex; in $d$ dimensions it involves seven irreducible pieces, \textit{i.e.}, one more than in 3 dimensions:
\begin{align}\label{decompFij}
\mathcal{F}_L^{ij} &= U^{(+2)}_{ijL} + \underset{L}{\text{STF}}\,
\underset{ij}{\text{STF}} 
\,\Bigl[ U^{(+1)}_{i | i_\ell j L-1} + \delta_{i i_\ell} U^{(0)}_{jL-1} + \delta_{i i_\ell} U^{(-1)}_{j |
	i_{\ell-1} L-2} + \delta_{i i_\ell} \delta_{j i_{\ell-1}}
U^{(-2)}_{L-2}  \nonumber\\ & \qquad \qquad \qquad \qquad \quad
 + W_{ij | i_{\ell}i_{\ell-1}L-2}\Bigr] + \delta_{ij} V_L\,.
\end{align}
As before, the vertical bars splitting the indices of the tensors $U^{(+1)}_{i | i_\ell j L-1}$, $U^{(-1)}_{j | i_{\ell-1} L-2}$ and $W_{ij | i_{\ell}i_{\ell-1}L-2}$, recall their non-trivial symmetries.  The penultimate piece, $W_{iji_{\ell}i_{\ell-1}L-2}$, is absent in 3 dimensions. Among these tensors, only $U^{(+2)}_{L i_{\ell+1} i_{\ell+2}}$, $U^{(0)}_L$, $U^{(-2)}_{L -2}$ and $V_L$ are STF in all their indices. We find
\begin{subequations}
	\begin{align}
U^{(+2)}_{L i_{\ell+1} i_{\ell+2}} &=  \mathcal{F}^{\langle i_{\ell+1}
	i_{\ell+2}}_{\phantom{\langle i_{\ell+1} i_{\ell+2}}L \rangle} \,, \\
U^{(+1)}_{i | i_{\ell+1} L} &= \frac{4\ell}{\ell+2}
\,\mathcal{P}^{(i_\ell\phantom{L-1}[i}_{\phantom{(i_\ell}L-1)\phantom{[i} i_{\ell+1}]} \,,\\ 
U^{(0)}_L &= \frac{2d\ell (2\ell + d -4)}{(d-2)(\ell+d-2)(2\ell+d)}
\,\underset{L}{\text{STF}}  \mathcal{F}^{\langle a i_\ell \rangle}_{~aL-1}  \,,\\
U^{(-1)}_{i | i_{\ell-1} L-2} &=
\frac{4(\ell-1)(2\ell+d-4)}{(\ell+d-2)(2\ell+d-2)} 
\,\mathcal{K}^{[i}_{\phantom{[i}i_{\ell-1}]L-2} \,,\\
U^{(-2)}_{L -2} &= \frac{\ell (\ell-1)
	(2\ell+d-6)}{(\ell+d-3)(\ell+d-4)(2\ell+d-2)}\,\mathcal{F}^{\langle ab
	\rangle}_{~abL-2} \,,\\
W_{ij | i_{\ell}i_{\ell-1}L-2} &=
\frac{4(\ell-1)}{\ell+1}
\,\mathcal{P}^{[j\phantom{i_{\ell}]}[i}_{\phantom{[j} i_{\ell}]  
	\phantom{[i} i_{\ell-1}] L-2} \,,\\
V_L &= \frac{1}{d} \,\mathcal{F}^{aa}_L \,. 
\end{align}\end{subequations}
Following~\eqref{TF1}, we have introduced, as convenient intermediates, the TF (but not symmetrized) parts $\mathcal{K}^{i}_{L-1} \equiv \underset{iL-1}{\text{TF}} \, \mathcal{F}^{\langle ai \rangle}_{aL-1}$ and $\mathcal{P}^{ij}_{L} \equiv \underset{ijL}{\text{TF}} \, \mathcal{F}^{ij}_L$, which are given explicitly by 
\begin{subequations}\label{TF2}
\begin{align}
\mathcal{K}^i_{L-1} &= \mathcal{F}^{\langle ia \rangle}_{~aL-1} 
- \frac{(\ell-1)(2\ell+d -6)}{(\ell+d-4)(2\ell+d-4)} \delta_{i\langle i_{\ell-1}}
\mathcal{F}^{\langle ab \rangle}_{L-2\rangle ab}
\, ,\\
\mathcal{P}^{ij}_{L} &= \mathcal{F}^{\langle ij \rangle}_{~L} - 
\underset{L}{\text{STF}}\,
\underset{ij}{\text{STF}} 
\,\Bigl[\delta_{i i_\ell} U^{(0)}_{jL-1} + \delta_{i i_\ell} \delta_{j i_{\ell-1}}
U^{(-2)}_{L-2} + \delta_{i i_\ell} U^{(-1)}_{j | i_{\ell-1}L-2} \Bigr] \, .
\end{align}\end{subequations}

Having established the irreducible decomposition of $\mathcal{F}_L^{\mu\nu}$, we must next deal (following the same steps as in~\cite{B98mult, BDEI05dr}) with the irreducible decomposition of the tensors $\mathcal{G}_L^{\mu}$ parametrizing the divergence of the multipole expansion~\eqref{Delta3}:
\begin{subequations}
\begin{align}
\partial_\nu\mathcal{M}(\Delta^{\mu\nu}) &= - \frac{4 G}{c^4}
\sum_{\ell=0}^{+\infty}
\frac{(-)^\ell}{\ell!}\,\partial_L\widetilde{\mathcal{G}}_L^{\mu} \,,\\
\text{with}\quad \widetilde{\mathcal{G}}_L^{\mu}(r,t) &=
\frac{\tilde{k}}{r^{d-2}}\int_1^{+\infty} \dd
z\,\gamma_{\frac{1-d}{2}}(z)\,\mathcal{G}_L^{\mu}(t-z r)\,.
\end{align}\end{subequations}
The function $\mathcal{G}_L^{\mu}$ is clearly not independent from $\mathcal{F}_L^{\mu\nu}$. It admits a closed form expression similar to~\eqref{calF0} in which an explicit factor $B$ comes out because of the differentiation of the regulator $\tilde{r}^B$, namely (the overdot means the time derivative):
\begin{subequations}
\begin{align}
\mathcal{G}_L^{\mu} &= -\ell\,\mathcal{F}_{L-1
	\rangle}^{\mu\langle i_\ell} + \frac{1}{c}\dot{\mathcal{F}}_L^{\mu 0} - \frac{1}{c^2(2\ell+d)}\ddot{\mathcal{F}}_{aL}^{a\mu} \label{divDelta3a}\\
&=\mathop{\mathrm{FP}}_{B=0}\int
\dd^d\mathbf{x}\,B \,\tilde{r}^{B} \,\frac{n_i}{r}\,
\hat{x}_L\,\overline{\tau}_{[\ell]}^{\mu i}(\mathbf{x},t)\,,\label{GLmu}
\end{align}\end{subequations}
where we recall the defining expressions~\eqref{exactform}--\eqref{PNform} of $\overline{\tau}_{[\ell]}^{\mu \nu}$. We decompose $\mathcal{G}_L^{\mu}$ in exactly the same way as in~\eqref{decompF00}--\eqref{decompF0i}, thus posing
\begin{subequations}\label{decompG}
\begin{align}
\mathcal{G}_L^{0} &= P_L \,,\\
\mathcal{G}_L^{i} &= Q^{(+)}_{iL} + Q^{(0)}_{i | \langle i_\ell L-1 \rangle} + \delta_{i \langle i_\ell} Q^{(-)}_{L-1 \rangle} \,.
\end{align}\end{subequations}
Notice that the relation~\eqref{divDelta3a} is equivalent to the following explicit constraints
\begin{subequations}\label{harmconstr}
\begin{align}
P_L &= \frac{1}{c} \dot{R}_L + -\ell \, T^{(+)}_{L} - \frac{(\ell + d - 2)}{(\ell + 1) (2\ell + d - 2) c^2} \ddot{T}^{(-)}_{L} \,,\\
Q^{(+)}_{L i_{\ell+1}} &= \frac{1}{c} \dot{T}^{(+)}_{L i_{\ell+1}} - \ell \, U^{(+2)}_{L i_{\ell+1}} - \frac{(d - 2) (\ell + d - 1) (2\ell + d + 2)}{2d (\ell + 1) (2\ell + d - 2) (2 \ell + d)c^2} \ddot{U}^{(0)}_{L i_{\ell+1}} \nonumber\\&- \frac{1}{(2\ell + d) c^2} \ddot{V}_{L i_{\ell+1}} \,,\\
Q^{(0)}_{i | i_\ell L-1} &= \frac{1}{c} \dot{T}^{(0)}_{i | i_\ell L-1} - \frac{\ell}{2} \ddot{U}^{(+1)}_{i | i_\ell L-1} - \frac{(\ell + d - 1)}{2 (\ell + 1 )(2 \ell + d - 2)} \ddot{U}^{(-1)}_{i | i_\ell L-1} \,,\\
Q^{(-)}_{L-1} &=	\frac{1}{c} \dot{T}^{(-)}_{L-1} - \frac{(d - 2) \ell}{2 d} U^{(0)}_{L-1} - \ell V_{L-1} - \frac{(\ell + d - 2)}{(\ell + 1) (2 \ell + d - 2) c^2} \ddot{U}^{(-2)}_{L-1} \,.
\end{align}\end{subequations}
Next, with those definitions in hands, we can pose
\begin{subequations}\label{vmunu}
\begin{align}
v^{00} &= -4G \biggl[- \hbox{$\int$} \Pt \,\dd t + \partial_a \bigg( \hbox{$\int$} \Pt_a \,\dd t + \hbox{$\int\!\!\int$} \Qpt_a \dd t - \frac{3d+1}{2d} \Qmt_a \bigg)\biggr]\, ,\\
v^{0i} &= - 4G \bigg[ -\hbox{$\int$} \Qpt_i \dd t
+ \frac{3d+1}{2d} \,\dQmt_i + \partial_a \bigg( \hbox{$\int$} \Qzt_{i|a} \dd t\bigg) - \sum_{\ell=2}^{+\infty}
\frac{(-)^\ell}{\ell!} \partial_{L-1} \Pt_{iL-1}\bigg]\, ,\\
v^{ij} &= -4G \biggl\{ \delta_{ij} \,\Qmt
+ \sum_{\ell= 2}^{+\infty} \frac{(-)^\ell}{\ell!} \biggl[ 2 \delta_{ij}
\partial_{L-1} \Qmt_{L-1} - 6 \partial_{L-2(i} \Qmt_{j)L-2} \\
& \qquad \quad  + \partial_{L-2} \biggl(
\dot{\Pt}_{ijL-2} 
-  \frac{7\ell+3d-6}{(\ell +1)(d+2\ell-2)} \ddQmt_{ijL-2}
+\ell  \Qpt_{ijL-2} \biggr) - 2 \partial_{L-1}
\Qzt_{(i|\underline{L-1}j)} \biggr]\biggr\}\,.\nonumber
\end{align}\end{subequations}
We remind that the overtilde means a monopolar homogeneous solution of the wave equation as defined by~\eqref{tilde}. In the last term, the underlined indices are excluded from the symmetrization operation, indicated by parenthesis $(\cdots)$. Finally, the integrals correspond to time anti-derivatives associated with GW losses: $\int\widetilde{P}\,\dd t \equiv \int_{-\infty}^t \dd t'\widetilde{P}(r,t')$, $\iint \widetilde{Q}\,\dd t \equiv \int_{-\infty}^t \dd t'\int_{-\infty}^{t'} \dd t'' \,\widetilde{Q}(r,t'')$.\footnote{We assume that the gravitational-wave source was stationary in the remote past. All the functions which need to be time integrated are zero in the past.}

\subsection{The irreducible source multipole moments in \boldmath $d$ \unboldmath dimensions} 

The role of the term $v^{\mu\nu}$ introduced in~\eqref{vmunu} is to ensure that 
\begin{align}\label{h1}
G h^{\mu\nu}_{1} \equiv \mathcal{M}(\Delta^{\mu\nu})- v^{\mu\nu}\,,
\end{align}
is at once a solution of the wave equation: $\Box h^{\mu\nu}_{1}=0$ (in $d$ dimensions) and divergence-free: $\partial_\nu h^{\mu\nu}_{1}=0$. This implies that $h^{\mu\nu}_{1}$ takes the form of a linearized solution (hence the subscript 1 and the factor $G$ inserted) of the relaxed field equations~\eqref{EE} in vacuum. The irreducible multipole moments can then be read off from that ``linearized'' solution~\cite{B98mult}.

Indeed, the general multipolar expansion decomposes as
\begin{subequations}\label{generalM}
\begin{align}
\mathcal{M}(h^{\mu\nu}) = G h^{\mu\nu}_{1} + u^{\mu\nu} + v^{\mu\nu}\,,\\ \text{where}~~u^{\mu\nu} \equiv \mathop{\mathrm{FP}}_{B=0}\,
\Box^{-1}_\mathrm{ret} \Bigl[ r^B \mathcal{M}(\Lambda^{\mu\nu})\Bigr]\,.
\end{align}\end{subequations}
In that formulation, the linearized metric $h^{\mu\nu}_{1}$, which is parametrized by the irreducible multipole moments ($I_L$, $J_{i\vert L}$ and $K_{ij\vert L}$ in $d$ dimensions), appears to be the ``seed'' of the multipolar-post-Minkowskian (MPM) iteration, \textit{i.e.}, 
\begin{equation}\label{generalMsum}
\mathcal{M}(h^{\mu\nu}) = \sum_{n=1}^{+\infty} G^n h^{\mu\nu}_{n}\,,
\end{equation}
where each of the non-linear coefficients $h_n^{\mu\nu} = u_n^{\mu\nu} + v_n^{\mu\nu}$ (for any $n\geqslant2$) is computed by induction, following exactly the MPM algorithm~\cite{BD86}.

The linearized metric $h^{\mu\nu}_{1}$ splits into a ``canonical'' metric $h^{\mu\nu}_{1\,\text{can}}$ having Thorne's form (see Eqs.~(8.12) in~\cite{Th80}), plus a linearized gauge transformation with gauge vector $\xi^\mu_1$:
\begin{align}\label{h1gauge}
h^{\mu\nu}_{1} =  h^{\mu\nu}_{1\,\text{can}} + \partial^\mu \xi^\nu_{1} +
\partial^\nu \xi^\mu_{1} - \eta^{\mu\nu} \partial_\lambda \xi^\lambda_{1}\,.
\end{align}
The above considerations lead us to the conclusion that, in $d$ dimensions, the canonical metric is parametrized by three irreducible moments $I_L$, $J_{i\vert L}$ and $K_{ij\vert L}$, as 
\begin{subequations}\label{h1can}
\begin{align}
h^{00}_{1\,\text{can}} &= - \frac{4}{c^2} \sum_{\ell=0}^{+\infty}
\frac{(-)^\ell}{\ell!} \partial_L \It_L \, ,\\
h^{0i}_{1\,\text{can}} &= \frac{4}{c^3} \sum_{\ell=0}^{+\infty}
\frac{(-)^\ell}{\ell!} \bigg[ \partial_{L-1} \dot{\It}_{iL-1}
+ \frac{\ell}{\ell+1} \partial_L \Jt_{i|L} \bigg] \, ,\\
h^{ij}_{1\,\text{can}} &= - \frac{4}{c^4} \sum_{\ell=0}^{+\infty}
\frac{(-)^\ell}{\ell!} \bigg[ \partial_{L-2} \ddot{\It}_{ijL-2}
+ \frac{2\ell}{\ell+1} \partial_{L-1} \dot{\Jt}_{(i|\underline{L-1}j)}
+ \frac{\ell-1}{\ell+1}\partial_L \Kt_{ij|L}\bigg] \,,
\end{align}\end{subequations}
and that these irreducible moments are given in terms of the previously used quantities by
\begin{subequations}\label{IJKL}
	\begin{align}
I_L &= R_L+ \frac{d}{d - 2} V_L - \frac{2 (d - 1)}{(\ell + 1) (d - 2) c}
      \dot{T}^{(-)}_L + \frac{(d - 1)}{(d - 2) (\ell + 1) (\ell + 2) c^2}
      \ddot{U}^{(-2)}_L \nonumber \\ &- \frac{2 (d- 3)}{(d - 2) (\ell + 1)} Q^{(-)}_L\,,\\
J_{i|L} &= - \frac{(\ell + 1)}{\ell c} T^{(0)}_{i | L} + \frac{1}{2 \ell c^2} \dot{U}^{(-1)}_{i | L}  \,,\\
K_{ij\vert L} &= \frac{\ell + 1}{\ell - 1}W_{ij | L}\,.
\end{align}\end{subequations}

Since we know the expressions~\eqref{calF0} and~\eqref{GLmu}, we can get our final explicit results for the moments. Posing 
\begin{align}\label{Sigma}
\overline{\Sigma} = \frac{2}{d-1}\,
\frac{(d-2)\overline{\tau}^{00}+\overline{\tau}^{ii}}{c^2}\,,\qquad
\overline{\Sigma}^i = \frac{\overline{\tau}^{i0}}{c}\,,\qquad
\overline{\Sigma}^{ij} = \overline{\tau}^{ij}\,,
\end{align}
where we recall that $\overline{\tau}^{\mu\nu}$ is the PN expansion of the pseudo-tensor $\tau^{\mu\nu}$, we obtain the mass-type moments as
\begin{align}\label{ILfinal}
I_L &= \frac{d-1}{2(d-2)}\mathop{\mathrm{FP}}_{B=0} \int
\dd^d\mathbf{x}\,\tilde{r}^B
\Biggl\{\hat{x}^L\,\overline{ \Sigma}_{[\ell]}-\frac{4(d+2\ell-2)}
{c^2(d+\ell-2)(d+2\ell)}\,\hat{x}^{iL}\,
\dot{\overline{\Sigma}}^{i}_{[\ell+1]}\nonumber\\
&\qquad\qquad\qquad~~ +\frac{2(d+2\ell-2)}
{c^4(d+\ell-1)(d+\ell-2)(d+2\ell+2)}
\,\hat{x}^{ijL}\,
\ddot{\overline{\Sigma}}^{ij}_{[\ell+2]}\nonumber\\
&\qquad\qquad\qquad~ -
\frac{4(d-3)(d+2\ell-2)}{c^2(d-1)(d+\ell-2)(d+2\ell)}
B \,\hat{x}^{iL}\,\frac{x^j}{r^2}
\,\overline{\Sigma}^{ij}_{[\ell+1]}
\Biggr\}\,,
\end{align}
in agreement with Eq.~(3.50) in~\cite{BDEI05dr}. Note that each of those terms can be computed more explicitly as PN expansions using~\eqref{PNform}. The last term in~\eqref{ILfinal} does not exist in 3 dimensions and plays no role in practical calculations of the mass moment up to the 4PN order. The mass moments are just STF in all their indices so that their symmetry is given by a symmetric Young tableau (with the multi-index $L=i_1\cdots i_\ell$)
\begin{align}\label{ILyoung}
I_L=\ytableausetup{boxsize=1.25em,textmode}
\ytableaushort{{\scriptsize $i_\ell$} {...} {\scriptsize $i_1$}}\,.
\end{align}
Next, we find the main theoretical output of this paper, which is the proper generalization of the current-type moments in $d$ dimensions, as
\begin{align}\label{JiLfinal}
J_{i|L} &= \mathop{\mathcal{A}}_{i i_{\ell}} \mathop{\text{FP}}_{B=0} \int
\dd^d \mathbf{x} 
\,\tilde{r}^{B} \Biggl\{-2
\biggl[ \hat{x}^L \,\overline{\Sigma}^{i}_{[\ell]} - 
\frac{\ell (2\ell 
	+d-4)}{(\ell+d-3)(2\ell+d-2)} \delta^{i\langle i_\ell} \hat{x}^{L-1\rangle
	a} \overline{\Sigma}^{a}_{[\ell]} \biggr] \\& + \frac{2 (2\ell +d-2)}{c^2(\ell+d-1)(2\ell+d)}
\biggl[ \hat{x}^{aL}\,\dot{\overline{\Sigma}}^{ia}_{[\ell+1]} - \frac{\ell (2\ell 
	+d-4)}{(\ell+d-3)(2\ell+d-2)} \delta^{i\langle i_\ell} \hat{x}^{L-1\rangle
	ab} \dot{\overline{\Sigma}}^{ab}_{[\ell+1]} \biggr]\Biggr\}\,.\nonumber
\end{align}
Here, $\mathcal{A}_{i i_{\ell}}$ means the anti-symmetrization with respect to the pair of indices $i i_{\ell}$ (with the factor $\frac{1}{2}$ included). Remark that the second terms inside the square brackets correspond to the $d$-dimensional trace of the first terms. Finally, as already mentioned, there exists in $d$ dimensions an additional type of irreducible multipole moment, which reads
\begin{align}\label{KijLfinal}
K_{ij|L} &= 4 \mathop{\mathcal{A}}_{j i_{\ell}} \mathop{\mathcal{A}}_{i i_{\ell-1}}
\mathop{\text{STF}}_{ij} \mathop{\text{STF}}_{L}  \mathop{\text{FP}}_{B=0} \int
\dd^d \mathbf{x} \,\tilde{r}^{B} \Biggl\{ \hat{x}^L
\,\overline{\Sigma}^{ij}_{[\ell]} \nonumber \\ & - \frac{2\ell (2 \ell + d - 4)
\Big[ (2(d - 2) \ell + d^2 -2 d + 4) \delta^{i i_\ell} \hat{x}^{L-1 a}
\,\overline{\Sigma}^{ja}_{[\ell]} + 4 (\ell - 1) \delta^{i i_\ell} \hat{x}^{L-2ja}
\,\overline{\Sigma}^{i_{\ell - 1} a}_{[\ell]} \Big]}{(d - 2) (\ell + d - 2) (2 \ell
+ d - 2) (2 \ell + d)} \nonumber \\ & \qquad + \frac{(\ell - 1) \ell (2 \ell + d - 4) (2 (d
- 2) \ell + d^2 - 4 d + 12)}{(d - 2) (\ell + d -3) (\ell + d - 2) (2 \ell + d - 2)
(2 \ell + d)}\delta^{j i_\ell} \delta^{i i_{\ell-1}} \hat{x}^{L-2ab}
\,\overline{\Sigma}^{ab}_{[\ell]} \nonumber \\ & \qquad + \frac{2 \ell (2 \ell + d -4)}{(d
-2) (\ell + d -2) (2 \ell + d)} \delta^{i i_\ell} \hat{x}^{L-1j}
\,\overline{\Sigma}^{kk}_{[\ell]} \Biggl\} \,.
\end{align}
The structure of this object is comparatively simpler than for $I_L$ and $J_{i|L}$ in the sense that it depends only on the tensorial piece $\overline{\Sigma}^{ij}_{[\ell]}$. The symmetries of the moments~\eqref{JiLfinal} and~\eqref{KijLfinal} are given by the mixed Young tableaux~\cite{james_1984, ma2004problems, Bekaert:2006py}
\begin{align}\label{JLKLyoung}
J_{i\vert L}=
\ytableaushort{{\scriptsize $i_\ell$} {\scriptsize
		$\,i_{\scalebox{0.6}{\text{$\ell\! -\! 1$}}}$} {...} {\scriptsize 
		$i_1$}, {\scriptsize $i$}}~\,,\qquad K_{ij\vert L} = \ytableaushort{{\scriptsize $i_\ell$} {\scriptsize
		$\,i_{\scalebox{0.6}{\text{$\ell\!- \! 1$}}}$} {\scriptsize
		$i_{\scalebox{0.6}{\text{$\ell\!- \! 2$}}}$} {...} {\scriptsize $i_1$},
	{\scriptsize $j$} 
	{\scriptsize $i$}}~\,,
\end{align}
respectively, with the convention that the indices are symmetrized over lines \textit{before} being antisymmetrized over columns. 

The tensor $K_{ij|L}$ looks unfamiliar because it actually vanishes in $3$ dimensions. This can be checked explicitly by expanding the expression~\eqref{KijLfinal} in an orthonormal triad. Another way to see this is to count the number of independent components of $K_{ij|L}$, which follows from the King's rule~\cite{king_1971} and is given by Eq.~\eqref{eq:nbcompP} in the Appendix. As it is proportional to $d-3$, there is no independent component in 3 dimensions. This is similar to what happens for the Weyl tensor, which does not exist either in 3 dimensions. In fact, the tensor $K_{ij | L}$ for $\ell=2$ has exactly the same trace-free property, symmetries, and number of independent components as the Weyl tensor.

It is convenient to introduce the following specific notation that allows reconstructing the symmetries of $J_{i|L}$ given $\ell+1$ indices:
\begin{equation}\label{SymJkji}
\underset{i|L}{\text{Sym}} \equiv \mathcal{A}_{i i_{\ell}} \underset{iL}{\text{TF}}\; \underset{L}{\text{STF}}.
\end{equation}
As an example, the first line of \eqref{JiLfinal} can be constructed, starting from $x^L\overline{\Sigma}^{i}_{[\ell]}$, by taking its STF part on the indices $L$, then removing the traces of the resulting object in $d$ dimensions and, finally, anti-symmetrizing on $\{i,i_\ell\}$. Thus, $J_{i\vert L}$ is trace-free, STF with respect to $L-1=i_{\ell-1}\cdots i_1$ and anti-symmetric with respect to the pair $i_\ell i$; on the other hand, $K_{ij\vert L}$ is trace-free, STF with respect to $L-2=i_{\ell-2}\cdots i_1$ and anti-symmetric with respect to both pairs $i_\ell j$ and $i_{\ell-1} i$. 

The gauge vector $\xi^\mu_{1}$ in~\eqref{h1gauge} can be dealt with in the same way as $h^{\mu\nu}_{1\,\text{can}}$. It admits the following irreducible decomposition:
\begin{subequations}
\begin{align}
& \xi^0_{1} = \frac{4}{c^3}\sum_{\ell=0}^{+\infty} \frac{(-)^\ell}{\ell!}\,\partial_L \Wt_L\, ,\\
& \xi^i_{1} =  -\frac{4}{c^4}\sum_{\ell=0}^{+\infty} \frac{(-)^\ell}{\ell!}\,\partial_{iL} \Xt_L
- \frac{4}{c^4}\sum_{\ell=1}^{+\infty} \frac{(-)^\ell}{\ell!}\,\biggl[ \partial_{L-1} \Yt_{L-1} + \partial_{L-1}
\Zt_{i|L-1}\biggr]\,,
\end{align}\end{subequations}
where the closed form expressions of the so-called gauge moments $\Wt_L$, $\Xt_L$, $\Yt_{L-1}$ (mass type) and $\Zt_{i|L-1}$ (current type) are similar to~\eqref{ILfinal} and~\eqref{JiLfinal}--\eqref{KijLfinal} but will not be needed henceforth. The gauge moments enter the signal at a relatively high PN order and can thus be computed at a low PN order, where one does not need to resort to dimensional regularisation at all. Therefore, we can just use here the 3-dimensional expressions of the gauge moments displayed in Eqs.~(5.17)--(5.20) of~\cite{B98mult}.

The ordinary STF mass-type moment in 3 dimensions, say $I^{[3]}_L$, is simply recovered as the limit $I_L^{[3]} = \lim_{d\to 3} I_L$ (after renormalization). In particular, we observe that this limit removes the last term in~\eqref{ILfinal}. For the ordinary STF current-type moment in 3 dimensions, $J^{[3]}_L$, we have
\begin{align}\label{lim3d}
	\lim_{d\to 3} J_{i\vert L} = \varepsilon_{ii_{\ell}a} \,J^{[3]}_{aL-1} \quad\Longleftrightarrow\quad J^{[3]}_{L} = \frac{1}{2}
	\,\varepsilon_{ab(i_\ell} \,\lim_{d\to 3}\, J_{\underline{a}|\underline{b} L-1)} \,,
	\end{align}
where $\varepsilon_{abi}$ is the usual Levi-Civita symbol in 3 dimensions (underlined indices being excluded from symmetrization). Notice that $J^{[3]}_L$, as recovered from~\eqref{lim3d}, not only is symmetric in its indices $L$ but is also automatically trace-free. Finally, as we said, the multipole moment $K_{ij\vert L}$ does not exist in 3 dimensions. 

We end this section by recalling, for completeness, the expressions of the STF moments in 3 dimensions, namely:
\begin{subequations}\label{moments3d}
	\begin{align}
	I^{[3]}_L(t) &= \mathop{\text{FP}}_{B=0} \int \dd^3\mathbf{x} \,\tilde{r}^B \int^1_{-1}
	\dd z \biggl[ \delta_\ell(z)\,\hat{x}_L \,\overline{\Sigma} -
	\frac{4(2\ell+1)}{c^2(\ell+1)(2\ell+3)} \delta_{\ell+1}(z)
	\,\hat{x}_{iL} \,\dot{\overline{\Sigma}}_i \\ & \qquad \qquad \qquad \qquad \qquad
	+ \frac{2(2\ell+1)}{c^4(\ell+1)(\ell+2)(2\ell+5)}
	\delta_{\ell+2}(z) \,\hat{x}_{ijL} \,\ddot{\overline{\Sigma}}_{ij} \biggr](t + z r/c) \,,\nonumber \\
	J^{[3]}_L(t) &= \mathop{\text{FP}}_{B=0} \int \dd^3\mathbf{x} \,\tilde{r}^B \int^1_{-1} \dd z
	\, \varepsilon_{ab \langle i_\ell} \biggl[ \delta_\ell(z) \,\hat{x}_{L-1
		\rangle a} \,\overline{\Sigma}_b \nonumber\\ & \qquad \qquad \qquad \qquad \qquad - \frac{2\ell+1}{c^2(\ell+2)(2\ell+3)}
	\,\delta_{\ell+1}(z) \,\hat{x}_{L-1 \rangle ac} \,\dot{\overline{\Sigma}}_{bc} \biggr](t + z r/c) \,,
	\end{align}\end{subequations}
where $\overline{\Sigma}=(\overline{\tau}^{00}+\overline{\tau}^{ii})/c^2$, $\overline{\Sigma}^{0i}=\overline{\tau}^{0i}/c$ and $\overline{\Sigma}^{ij}=\overline{\tau}^{ij}$ (in 3 dimensions), whereas $\delta_\ell (z)$ is the $3d$ limit of the counterpart quantity introduced in~\eqref{exactform}, so that
\begin{subequations}\label{intdelta3d}
\begin{align}
\int^1_{-1} \dd z \,\delta_\ell(z) \,\overline{\Sigma}(\mathbf{x}, t+z r/c) &= \sum_{k=0}^{+\infty}\frac{(2\ell+1)!!}{2^k k!(2\ell+2k+1)!!} \biggl(\frac{r}{c}\frac{\partial}{\partial t}\biggr)^{\!2k} \!\overline{\Sigma}(\mathbf{x}, t)\,,\\
\delta_\ell (z) &\equiv \frac{(2\ell+1)!!}{2^{\ell+1} \ell!}(1-z^2)^\ell\,.
\end{align}\end{subequations}

\section{The current quadrupole moment of compact binaries} 
\label{sec:currentquad3PN}

In this section, we compute $J_{k|ji}$ in $d$-dimensions for (non-spinning) compact binaries up to the 3PN order using (\ref{JiLfinal}) for $\ell=2$. Remember that $J_{k|ji}$ is merely the dual of the physical current quadrupole moment $J_{ij}$ but remark also that it is not symmetric on $\{i,j\}$. We employ essentially the same set of techniques as for the computation of the mass quadrupole moment at the 4PN order~\cite{MHLMFB20}; all calculations are performed with \textit{Mathematica} supplemented with the \textit{xAct} library~\cite{xtensor}. Here are the salient points of the method:
\begin{itemize}
\item The PN metric in harmonic coordinates is expressed in terms of elementary potentials satisfying a set of ordinary nested wave equations.
\item We express $J_{k|ji}$ in terms of integrals of potentials by inserting the sources~\eqref{Sigma} into~\eqref{JiLfinal}. We then simplify the integrand as much as possible using integration by part, together with a set of relations between potentials (see Sec.~III C in~\cite{MHLMFB20}).
\item The resulting expression for $J_{k|ji}$ is divided into three types of terms: compact support terms, non-compact support terms and surface terms at spatial infinity (when $r\to+\infty$ with fixed $t$). 
\item Compact support terms are calculated directly in $d$ dimensions using the $d$-dimensional regularized potentials at the source points $1$ and $2$ derived in~\cite{MHLMFB20}.
\item Non-compact support terms are first computed in 3 dimensions; the difference between dimensional and Hadamard regularisations in the UV is obtained in a second stage using Eq. (4.8) of~\cite{MHLMFB20}, and added to the 3-dimensional result. 
\item For one particular non-compact support term (involving the non-linear potential $\hat{Y_i}$ defined by~(A4h) in~\cite{MHLMFB20}), we employ the method of super-potentials (see also more details below).
\item All derivatives of potentials are understood as Schwartz distributional derivatives~\cite{Schwartz}, \textit{i.e.}, we apply the Gel'fand-Shilov formula in $d$ dimensions~\cite{gelfand} (see Sec.~IV C in~\cite{MHLMFB20}).
\item However, the IR divergences are treated using the Hadamard partie finie procedure with regulator $\tilde{r}^B$ in 3 dimensions (see Ref.~\cite{DDR_IR} for a justification of this point).
\item The result for $J_{k|ji}$ after integration contains a pole in $\tfrac{1}{d-3}$: the last step consists in applying a shift, given in Eq.~(B1) of~\cite{MHLMFB20}, on the bodies' positions; when substituting the renormalized position variables to the bare ones, the pole in $J_{k|ji}$ cancels out.
\item Finally, we can take the limit $d\rightarrow 3$ using~\eqref{lim3d} in order to get the renormalized current quadrupole moment $J_{ij}$ displayed in Eq.~\eqref{J2CMstruct}.
\end{itemize}
The investigation of the current moment in $d$ dimensions in Sec.~\ref{sec:currentmult} is essential to treat the UV divergences with dimensional regularisation, which incidentally is the only way to give a full meaning to distributional derivatives. However, our calculation of the current quadrupole limited to 3PN order does not require corrections from dimensional regularisation for the IR divergences. We thus treat those in the source moment with the Hadamard partie finie regularisation and add the non-linear corrections due to tails and tails-of-tails as computed in 3 dimensions. We have shown that the IR corrections in the source moment computed in pure dimensional regularisation are cancelled by corresponding UV divergences coming from the tails-of-tails in $d$ dimensions, yielding the equivalence of Hadamard partie finie and dimensional regularisation for the IR divergencies at 3PN order~\cite{DDR_IR}.

Let us illustrate one computational aspect concerning the use of ``super-potentials''. One non-compact support term to be computed in the current quadrupole moment $J_{k|ji}$ reads
\begin{align}\label{JijY}
J_{k|ji}^{\hat{Y}} \equiv -\frac{4}{\pi G c^6}\frac{d-1}{d-2}\, \underset{k|ji}{\text{Sym}}\mathop{\text{FP}}_{B=0} \int \dd^d\mathbf{x} \,\tilde{r}^B x^{ij} \,\partial_k \hat{Y}_a\,\partial_a V\,,
\end{align}
where Sym is defined in \eqref{SymJkji}. It involves the difficult non-linear potential $\hat{Y}_a$ obeying a Poisson equation $\Delta \hat{Y}_a = \hat{S}_a$ to leading order. While the source term $\hat{S}_a$ is known and relatively easy to manage, no closed-form expression is available for the potential $\hat{Y}_a$ itself, even in 3 dimensions. On the other hand, $V$ is a simple linear potential with compact support: $\Delta V = -4\pi G \sigma$.\footnote{See Appendix~A in~\cite{MHLMFB20} for a compendium of formulas and complete definitions of potentials.} By means of an integration by parts taking advantage of the symmetries of $J_{k|ji}$ (the all-integrated term being zero by analytic continuation in $B$), we transform this term into
\begin{align}\label{JijY2}
J_{k|ji}^{\hat{Y}} =\frac{4}{\pi G c^6}\frac{d-1}{d-2} \,\underset{k|ji}{\text{Sym}}\mathop{\text{FP}}_{B=0} \int \dd^d\mathbf{x} \,\tilde{r}^B \hat{x}^{ij}\,\hat{Y}_a \,\partial_{ka}V \,. 
\end{align}
The idea behind introducing super-potentials is that the solution of the equation $\Delta \Psi_{ij}^{\partial_{ka} V} = \hat{x}_{ij}\,\partial_{ka} V$ (where $\hat{x}_{ij}\equiv x_{\langle ij\rangle}$ is STF) is known in analytic closed-form in terms of the super-potentials of $V$, namely the Poisson-like potentials $V_{2k}$ satisfying the hierarchy of equations $\Delta V_{2k+2} = V_{2k}$, together with $V_0\equiv V$. This solution,
\begin{align}\label{Psija}
\Psi_{ij}^{\partial_{ka} V} = x_{\langle ij\rangle}\,\partial_{ka} V_2 - 4 x_{\langle i}\partial_{j\rangle ka} V_4 + 8 \partial_{\langle ij\rangle ka} V_6\,,
\end{align}
is valid in the sense of distributions and its expression holds in any $d$ dimensions.\footnote{The conditions under which this solution is unique have been investigated in Sec.~IVB of~\cite{BFW14b}.} Thanks to it, we can transform the term~\eqref{JijY2} after further integrations by parts into the more tractable form
\begin{align}\label{JijY3}
J_{k|ji}^{\hat{Y}} = \frac{4}{\pi G c^6}\frac{d-1}{d-2} \,\underset{k|ji}{\text{Sym}}\mathop{\text{FP}}_{B=0} \int \dd^d\mathbf{x} \,\tilde{r}^B\left[ \Psi_{ij}^{\partial_{k a} V} \hat{S}_a +\partial_b\left(\partial_b\Psi_{ij}^{\partial_{k a} V}\hat{Y}_a - \Psi_{ij}^{\partial_{k a} V}\partial_b\hat{Y}_a\right)\right]\,. 
\end{align}
Here, the first term is straightforwardly computed because we know the source $\hat{S}_a$ of the potential $\hat{Y}_a$. As for the second term, which is a surface term, it can be computed directly in 3 dimensions since it does not involve UV divergences. It depends only on the expansions of $\hat{Y}_a$ and the superpotential at spatial infinity, when $r\to +\infty$. The expansion of $\hat{Y}_a$ can be determined directly from the source term $\hat{S}_a$, without having to control $\hat{Y}_a$ all-over the space. As a check of the result, we have also calculated $J_{k|ji}^{\hat{Y}}$ using the alternative form
\begin{align}\label{JijY4}
  J_{k|ji}^{\hat{Y}} &= -\frac{4}{\pi G c^6}\frac{d-1}{d-2} \,\underset{k|ji}{\text{Sym}}\mathop{\text{FP}}_{B=0} \int \dd^d\mathbf{x} \,\tilde{r}^B\Big\{ 4 x^i \partial_a V_4 \partial_{jk} \hat{S}_a + x^{ij} \partial_a V_2 \partial_k \hat{S}_a \\ & \qquad \qquad +
  \partial_l \Big[ x^{ij} \partial_{al} V_2 \partial_k \hat{Y}_a - \partial_a V_2 \partial_l (x^{ij} \partial_k \hat{Y}_a) + 4 x^i \partial_{al} V_4 \partial_{kj} \hat{Y}_a - 4 \partial_a V_4 \partial_l (x^i \partial_{kj} \hat{Y}_a) \Big]  \Big\}\,,\nonumber
\end{align}
which is derived directly by substituting $\Delta \hat{Y}_a$ to $\hat{S}_a$, expanding the derivatives by means of the Leibniz rule, and resorting again to the properties of the symmetry operator $\text{Sym}_{k|ji}$.

The final expression of the 3PN current quadrupole in a general frame is obtained, as mentioned above, by applying the UV shift, taking the 3 dimensional limit and, finally, using~\eqref{lim3d} to come back to the usually looking current moment $J_{ij}$. We have checked that the symmetry operations \eqref{SymJkji} commute with the computation of the difference between the dimensional and Hadamard regularisations. The expression of the final current quadrupole is quite long; so we present it only in the center-of-mass (CM) frame.\footnote{Notation is as follows: The relative position and velocity of the two particles (in harmonic coordinates) are $x^i =y_1^i - y_2^i$, $v^i = dx^i/dt = v_1^i - v_2^i$; the distance between the two particles $r=\vert\bm{x}\vert$; the masses $m_1$ and $m_2$, the total mass $m=m_1+m_2$, the symmetric mass ratio $\nu=m_1m_2/m^2$ and the mass difference ratio $\Delta=(m_1-m_2)/m$; finally, we pose $L^i\equiv\varepsilon^{iab}x^a v^b$ and $(vx)=v^i x^i$.} For general orbits (bound or unbound), we obtain
\begin{align}\label{J2CMstruct}
J_{ij} = - \nu m \Delta\left[ A \,L^{\langle i} x^{j\rangle} + B \,\frac{(vx)}{c^2}\,L^{\langle i} v^{j\rangle} + C \,\frac{G m}{c^3}\,L^{\langle i} v^{j\rangle} \right] + \mathcal{O}\left(\frac{1}{c^7}\right)\,.
\end{align}
The coefficients $A$ and $B$ describe the conservative effects up to the 3PN order; $A$ also includes a 2.5PN dissipative term, while the coefficient $C$ is purely dissipative. We have 
\begin{subequations}\label{J2CMABC}
\begin{align}
A &= 1 + \frac{1}{c^{2}}\biggl[\Bigl( \frac{13}{28} - \frac{17}{7} \nu \Bigr) v^{2} + \frac{G m}{r} \Bigl( \frac{27}{14} +  \frac{15}{7} \nu \Bigr)\biggr] \nonumber\\
&~ + \frac{1}{c^{4}}\biggl[\Bigl( \frac{29}{84} - \frac{11}{3} \nu +  \frac{505}{56} \nu^2\Bigr) v^{4} + \frac{G m}{r^3} \Bigl( - \frac{5}{252} - \frac{241}{252} \nu - \frac{335}{84} \nu^2\Bigr) (v x)^2 \nonumber\\
&~\qquad + \frac{G m}{r} \Bigl( \frac{671}{252} - \frac{1297}{126} \nu - \frac{121}{12} \nu^2\Bigr) v^{2} + \frac{G^2 m^2}{r^2} \Bigl( - \frac{43}{252} - \frac{1543}{126} \nu +  \frac{293}{84} \nu^2\Bigr)\biggr] \nonumber\\
&~ + \frac{1}{c^5}\biggl[-\frac{62}{7} \frac{G^2 m^2}{r^3}\nu (vx) \biggr] \nonumber \\
&~ + \frac{1}{c^{6}}\Biggl[\frac{G m}{r^3} \Bigl( \frac{109}{11088} - \frac{2687}{1232} \nu - \frac{13507}{11088} \nu^2 +  \frac{31387}{924} \nu^3\Bigr) (v x)^2 v^{2} \nonumber\\
&~\qquad + \frac{G^2 m^2}{r^2} \Bigl( \frac{166931}{27720} - \frac{273443}{5544} \nu +  \frac{9283}{264} \nu^2 - \frac{145669}{5544} \nu^3\Bigr) v^{2} \nonumber\\
&~\qquad + \Bigl( \frac{709}{2464} - \frac{35059}{7392} \nu +  \frac{179615}{7392} \nu^2 - \frac{73517}{1848} \nu^3\Bigr) v^{6} \nonumber\\
&~\qquad + \frac{G m}{r^5} \Bigl( - \frac{67}{2464} +  \frac{1667}{2464} \nu +  \frac{7813}{7392} \nu^2 - \frac{3137}{462} \nu^3\Bigr) (v x)^4 \nonumber\\
&~\qquad + \frac{G m}{r} \Bigl( \frac{74083}{22176} - \frac{201835}{7392} \nu +  \frac{803783}{22176} \nu^2 +  \frac{25621}{462} \nu^3\Bigr) v^{4} \nonumber\\
&~\qquad + \frac{G^2 m^2}{r^4} \Bigl( - \frac{191}{560} - \frac{1909}{336} \nu - \frac{9073}{144} \nu^2 - \frac{13169}{504} \nu^3\Bigr) (v x)^2 \nonumber\\
&~\qquad + \frac{G^3 m^3}{r^3} \Bigl(  \frac{23443}{7425} +  \frac{79067}{9240} \nu +  \frac{123}{128} \pi^2 \nu - \frac{120185}{5544} \nu^2 +  \frac{85031}{16632} \nu^3 \nonumber\\
&~\qquad\qquad 
- \frac{214}{105} \ln\Bigl(\frac{r}{r_{0}}\Bigr) - 22 \nu \ln\Bigl(\frac{r}{r'_{0}}\Bigr) \Bigl) \Biggr]\,,\\
B &= \frac{5}{28} - \frac{5}{14} \nu \nonumber\\
&~+ \frac{1}{c^{2}}\biggl[\Bigl( \frac{25}{168} - \frac{25}{24} \nu +  \frac{25}{14} \nu^2\Bigr) v^{2} + \frac{G m}{r} \Bigl( \frac{103}{63} +  \frac{337}{126} \nu - \frac{173}{84} \nu^2\Bigr) \biggr] \nonumber\\
&~ + \frac{1}{c^{4}}\biggl[\Bigl( \frac{305}{2464} - \frac{11815}{7392} \nu +  \frac{49025}{7392} \nu^2 - \frac{32905}{3696} \nu^3\Bigr) v^{4} \nonumber\\
&~\qquad + \frac{G m}{r^3} \Bigl( \frac{1}{396} - \frac{3053}{1848} \nu +  \frac{14605}{5544} \nu^2 +  \frac{967}{462} \nu^3\Bigr) (v x)^2 \nonumber\\
&~\qquad + \frac{G m}{r} \Bigl( \frac{193}{99} - \frac{6095}{1848} \nu - \frac{17783}{1386} \nu^2 +  \frac{941}{66} \nu^3\Bigr) v^{2} \nonumber\\
&~\qquad + \frac{G^2 m^2}{r^2} \Bigl( \frac{164023}{18480} +  \frac{98485}{3696} \nu +  \frac{194041}{11088} \nu^2 - \frac{41215}{5544} \nu^3\Bigr) \biggr]\,,\\
C &= \frac{4\nu}{5 c^2}\biggl(\frac{54}{7}\frac{G m}{r} - v^{2}\biggr)\,.
\end{align}
\end{subequations}
At the occasion of this calculation, we have also re-computed the mass-type octupole moment $I_{ijk}$ at the 3PN order and confirm our previous result (in a general frame), which was given in the CM frame by Eqs.~(4.10) of~\cite{FBI15}.

Note the dependence on two regularisation scales $r_0$ and ${r'}_0$ through logarithms in the coefficient $A$. The scale $r_0$, which enters a logarithm with numerical prefactor coefficient $\beta_J=-\frac{214}{105}$, is due to the IR divergences of the multipole moment. This constant will be found to properly cancel out in the final expression of the GW mode $h^{21}$ in Sec.~\ref{sec:mode21}. In effective field theory (EFT), the cancellation of $r_0$ is ruled by the renormalization group equations and $\beta_J$ is the associated beta function coefficient.\footnote{Note that $\beta_J$ happens to be the same as $\beta_I=-\frac{214}{105}$, which is associated with the renormalization of the mass-type quadrupole moment~\cite{GRoss10}.} In the traditional PN approach adopted here, the constant $r_0$ is canceled by the same constant that is present in the tail-of-tail correction of the radiative current quadrupole [see Eq.~\eqref{Vij}]. As for ${r'}_0$, it is a UV scale and it also properly cancels out when we express the orbital separation $r$ of the particles in terms of the invariant orbital frequency $\omega$ or, equivalently, the PN parameter $x$ [see Eq.~\eqref{H21final}]. Following the discussion in Sec.~VI (footnote 10) of~\cite{MHLMFB20}, we identify for simplicity the two \textit{a priori} independent UV scales associated with the two particles, respectively, \textit{i.e.}, we set ${r'}_0\equiv{r'}_1={r'}_2$.

In the case of quasi-circular orbits, we get [with $\gamma \equiv Gm/(rc^2)$]
\begin{subequations}\label{J2circ}
\begin{align}
J_{ij} &= - \nu m \Delta\left[ A_\text{circ} \,L^{\langle i} x^{j\rangle} + C_\text{circ} \,\frac{G m}{c^3}\,L^{\langle i} v^{j\rangle} \right] + \mathcal{O}\left(\frac{1}{c^7}\right)\,,\\
A_\text{circ} &= 1 +\gamma \left(\frac{67}{28}-\frac{2}{7}\nu \right)+\gamma^2\left(\frac{13}{9} -\frac{4651}{252}\nu -\frac{\nu^2}{168} \right) \nonumber\\ & ~+ \gamma^3\left(\frac{2301023}{415800} - \frac{214}{105}\ln\left(\frac{r}{r_0}\right)\right. \nonumber\\&\qquad\qquad\left. + \left[ - \frac{243853}{9240} + \frac{123}{128}\pi^2 - 22 \ln\left(\frac{r}{r'_0}\right)\right]\nu + \frac{44995}{5544}\nu^2 + \frac{599}{16632}\nu^3\right)\,,\\
C_\text{circ} &= \frac{188}{35} \nu \,\gamma\,.
\end{align}\end{subequations}

\section{The gravitational-wave mode \boldmath $h_{21}$ \unboldmath at 3PN order} 
\label{sec:mode21}

Let us now review the expression of the \textit{radiative} current quadrupole moment $V_{ij}$ in terms of the source moment $J_{ij}$. The radiative moments parametrize the asymptotic waveform to leading order $1/R$ in the distance, within the class of radiative coordinates $(T, R)$ such that $u \equiv T - R/c$ is a null coordinate, or becomes asymptotically null in the limit $R\to +\infty$. In terms of the harmonic coordinates  $(t, r)$, we have
\begin{equation}\label{TRtr}
u = t - \frac{r}{c} - \frac{2 G M}{c^3}\ln\left(\frac{r}{c b}\right) + \mathcal{O}\left(\frac{1}{r}\right)\,,
\end{equation}
where $M$ is the conserved total mass of the source and $b$ an arbitrary constant time scale (independent from $r_0$). Here, we are interested in the radiative current quadrupole moment, linking to the source moment through a series of non-linear corrections. We first relate $V_{ij}$ to the so-called \textit{canonical} current quadrupole moment $S_{ij}$, which constitutes a useful intermediate definition. To 3PN order, we have
\begin{align}\label{Vij}
V_{ij} (u) &=
S^{(2)}_{ij} (u) \nonumber\\ &+ \frac{G M}{c^3} \int_{-\infty}^{u} \dd \tau \left[ 2 \ln
\left(\frac{u-\tau}{2b}\right)+\frac{7}{3} \right] S^{(4)}_{ij}
(\tau)\nonumber\\ &+ \frac{G}{7\,c^{5}}\Biggl\{4S^{(2)}_{a\langle
	i}M^{(3)}_{j\rangle a}+8M^{(2)}_{a\langle i}S^{(3)}_{j\rangle a}
+17S^{(1)}_{a\langle i}M^{(4)}_{j\rangle a}-3M^{(1)}_{a\langle
	i}S^{(4)}_{j\rangle a}+9S_{a\langle i}M^{(5)}_{j\rangle a}\nonumber\\
&\qquad\quad-3M_{a\langle i}S^{(5)}_{j\rangle
	a}-\frac{1}{4}S_{a}M^{(5)}_{ija}-7\varepsilon_{ab\langle
	i}S_{a}S^{(4)}_{j\rangle b} +\frac{1}{2}\varepsilon_{ac\langle
	i}\biggl[3M^{(3)}_{ab}M^{(3)}_{j\rangle bc} +\frac{353}{24}M^{(2)}_{j\rangle
	bc}M^{(4)}_{ab}\nonumber\\
&\qquad\quad-\frac{5}{12}M^{(2)}_{ab}M^{(4)}_{j\rangle
	bc}+\frac{113}{8}M^{(1)}_{j\rangle bc}M^{(5)}_{ab}
-\frac{3}{8}M^{(1)}_{ab}M^{(5)}_{j\rangle bc}+\frac{15}{4}M_{j\rangle
	bc}M^{(6)}_{ab} +\frac{3}{8}M_{ab}M^{(6)}_{j\rangle
	bc}\biggr]\Biggr\}\nonumber\\
& + \frac{G^2 M^2}{c^6} \int_{-\infty}^{u} \dd \tau \bigg[2 \ln^2
\bigg(\frac{u-\tau}{2b} \bigg) \nonumber \\ & \qquad\quad + \frac{14}{3} \ln
\bigg(\frac{u-\tau}{2b} \bigg) -
\frac{214}{105} \ln \bigg(\frac{c(u-\tau)}{2 r_0}\bigg) -
\frac{26254}{11025}\bigg] S_{ij}^{(5)}(\tau) + \,\mathcal{O}\left(\frac{1}{c^7}\right)\,.
\end{align}
At this accuracy level, the right-hand side contains the dominant quadratic tail term at 1.5PN order~\cite{BD92}, a number of instantaneous corrections at 2.5PN order involving the canonical moments $M_L$, $S_L$~\cite{BFIS08}, and the cubic tail-of-tail term at 3PN order~\cite{FBI15}. Note the presence of the IR scale $r_0$ in one of the logarithms of the tail-of-tail term in~\eqref{Vij}, with the same coefficient $\beta_J=-\frac{214}{105}$ as in the source moment~\eqref{J2CMstruct}, which shows that the constant $r_0$ will finally drop from the radiative moment. Next, the canonical moment $S_{ij}$ is related to the source moments $I_L$, $J_L$ and to the so-called gauge moments $W_L$, $X_L$, $Y_L$, $Z_L$ by~\cite{BFIS08}
\begin{align}
	S_{ij} &= J_{ij} +\frac{2G}{c^5}\biggl[\varepsilon_{ab\langle i}\left(-I_{j\rangle b}^{(3)}W_{a}-2I_{j\rangle b}Y_{a}^{(2)}+I_{j\rangle b}^{(1)}Y_{a}^{(1)}\right)+3J_{\langle i}Y_{j\rangle}^{(1)}-2J_{ij}^{(1)}W^{(1)}\biggr]
	+ \mathcal{O}\left(\frac{1}{c^7}\right)\,.
\end{align}
For all moments in~\eqref{Vij} but in the first term $S^{(2)}_{ij} (u)$, we can identify $M_L$, $S_L$ with $I_L$, $J_L$ at the 3PN approximation. The required gauge moments can be found in Eqs.~(5.7) of~\cite{FMBI12}.

With our control of the radiative current moment $V_{ij}$, we are in the position to compute the mode $h^{21}$ at the 3PN order. The modes are defined from the $+$ and $\times$ polarization waveforms as
\begin{align}
h \equiv h_{+} - \di h_{\times} = \sum^{+\infty}_{\ell=2}\sum^{\ell}_{m=-\ell} h^{\ell
m} \,Y^{\ell m}_{-2}(\Theta,\Phi)\,,
\end{align}
where the spin-weighted spherical harmonics of weight $-2$ are functions of the spherical angles $(\Theta,\Phi)$ defining the direction of propagation.\footnote{We adopt the same conventions as in Refs.~\cite{BFIS08, FMBI12}. The spin-weighted spherical harmonics are given by~(2.4) in~\cite{BFIS08}, and Fig.~1 in~\cite{FMBI12} specifies our convention for the polarization vectors.} For planar binaries, which are either non-spinning or with spins aligned or anti-aligned with the orbital angular momentum, we know that there is a clean separation of modes between mass-type and current-type contributions (see~\cite{K07} and Sec. III B in~\cite{FMBI12}). In particular, the ``current'' modes are entirely determined by the current radiative moments when $\ell+m$ is an odd integer,
\begin{align}
	h^{\ell m} &= \frac{\di}{\sqrt{2}}\frac{G}{R \,c^{\ell +3}} \,V^{\ell m}
	~\qquad\text{(for $\ell+m$ odd)} \,,
\end{align} 
hence the only relevant mode for the current quadrupole is $h^{21}$ (recall that $h^{2,-1} = \overline{h^{21}}$ in the planar case). The current moment in non-STF spin-weighted guise is given in terms of the STF version $V_L$ by
\begin{align}
V^{\ell m} &= -\frac{8}{\ell!}\,\sqrt{\frac{\ell(\ell+2)}{2(\ell+1)(\ell-1)}}\,\alpha_L^{\ell m}\,V_L\,,
\end{align}
where the STF tensorial coefficient $\alpha_L^{\ell m} \equiv \int \dd \Omega\,\hat{N}_L\,\overline{Y}^{\,\ell m}$ is defined from the ordinary spherical harmonics $Y^{\ell m}$ (or in fact its complex conjugate $\overline{Y}^{\,\ell m}$). In practice, we use the convenient orthonormal triad $(\boldsymbol{n},\boldsymbol{\lambda},\boldsymbol{l})$ where $\boldsymbol{n}= \boldsymbol{x}/|\boldsymbol{x}|$, $\boldsymbol{l}=\boldsymbol{L}/|\boldsymbol{L}|$, while $\boldsymbol{\lambda}$ completes the triad with right-handed orientation. We also define $\mathfrak{m}=(\boldsymbol{n}+ \di \boldsymbol{\lambda})/\sqrt{2}$, as well as its value $\mathfrak{m}_0$ at some reference time $t_0$, to obtain the explicit expression of $\alpha^{\ell m}_L$:\footnote{We have $\mathcal{Y}_L^{\ell m}=\frac{(2\ell+1)!!}{4\pi l!}\,\overline{\alpha}_L^{\ell m}$ in the alternative definition used in~\cite{Th80, K07}.}
\begin{equation}
\alpha^{\ell m}_L = \dfrac{\sqrt{4\pi}(-\sqrt{2})^m\ell!}{\sqrt{(2\ell+1)(\ell+m)!(\ell-m)!}}\,\overline{\mathfrak{m}}_0^{\langle M}l^{L-M\rangle}.
\end{equation}
In the tail and tail-of-tail terms at 1.5PN and 3PN orders, there appears the
total Arnowitt-Deser-Misner (ADM) mass $M$, which therefore needs to be
computed with 1PN precision. It is convenient, following~\cite{BIWW96,ABIQ04},
to perform a change of phase variable, from the actual orbital phase $\phi$ to
the new variable
\begin{equation}\label{psi}
\psi \equiv \phi - \frac{2G M \omega}{c^3} \ln\left(\frac{\omega}{\omega_0}\right)\,.
\end{equation}
The constant $\omega_0$, equivalent to $b$ in~\eqref{TRtr}, is conveniently defined by $\omega_0 = \frac{1}{4 b}\mathrm{exp}[\frac{11}{12}-\gamma_\text{E}]$ with $\gamma_\text{E}$ the Euler constant. The main advantage of the new phase variable~\eqref{psi} is that it minimizes the occurence of logarithms due to tails in the waveform and modes. Notice, however, that the use of either $\phi$ or $\psi$ is equivalent for the present 3PN level of accuracy. Indeed, the correction term in~\eqref{psi} is of order 1.5PN, which means that the effect as seen as a correction to the phase evolution is actually of order 4PN with respect to the leading order in the phase provided by the usual quadrupole formula. A similar variable is also introduced in black-hole perturbation theory~\cite{Sasa94,TSasa94,TTS96}. 

The modes are thus defined with respect to the phase variable~\eqref{psi} as
\begin{align}
	h^{\ell m} = \frac{2 G \,m \,\nu \,x}{R \,c^2} \,
	\sqrt{\frac{16\pi}{5}}\,\hat{H}^{\ell m}\,e^{-i m \psi}\,.
\end{align}
Beware that the symbol $m$ in the first factor of the right-hand side denotes the total mass $m=m_1+m_2$, whereas its two other occurrences refer to an integral label.
The final result for the 3PN mode 21 (corresponding in fact to the 3.5PN accurate waveform) is expressed with the usual gauge invariant PN parameter (with $\omega=\dot{\phi}$)
\begin{align}\label{x}
x\equiv\biggl(\frac{G m \omega}{c^3}\biggr)^{2/3}\,.
\end{align}
As already announced, the regularisation constants $r_0$, ${r'}_0$ and the gauge constant $b$ disappear from the end result, and we find, extending~(9.4b) in~\cite{BFIS08}:
\begin{align}\label{H21final}
\hat{H}^{21}
&=\frac{\di}{3} \,\Delta \Biggl[x^{1/2}+x^{3/2} \left(-\frac{17}{28}+\frac{5
	\nu }{7}\right)+x^2 \left(\pi +\di \biggl[-\frac{1}{2}-2 \ln 2\biggr]\right)
\\ &\qquad\quad +x^{5/2} \left(-\frac{43}{126}-\frac{509 \nu }{126}+\frac{79
	\nu^2}{168}\right)\nonumber \\ &\qquad\quad +x^3 \bigg(\pi\biggl[-\frac{17}{28}+\frac{3\nu }{14}\biggr] + \di \biggl[\frac{17}{56}+\nu \left(-\frac{353}{28}-\frac{3}{7}\ln
2\right)+\frac{17}{14}\ln 2\biggr]\bigg)\nonumber \\ &\qquad\quad + x^{7/2}\biggl(\frac{15223771}{1455300}+\frac{\pi^2}{6}-\frac{214}{105}\gamma_\text{E}-\frac{107}{105}\ln(4x)-\ln 2-2(\ln 2)^2\nonumber \\ &\qquad\quad\quad+\nu\biggl[-\frac{102119}{2376}+\frac{205}{128}\pi^2\biggr]-\frac{4211}{8316}\nu^2+\frac{2263}{8316}\nu^3+\di \pi\biggl[\frac{109}{210}-2\ln 2\biggr]\biggr)\Biggr] 
+ \mathcal{O}\left(\frac{1}{c^8}\right)\,.\nonumber
\end{align}
The perturbative limit $\nu\to 0$ is in perfect agreement with the result of black-hole perturbation theory~\cite{Sasa94,TSasa94,TTS96}, as one may check with the mode provided in this limit in the Appendix B in~\cite{TSasa94}.\footnote{For the comparison, note that the phase variable used in the Appendix B of~\cite{TSasa94} is related to ours by $\psi_\text{TS}=\psi+\pi/2+2x^{3/2}[\ln 2-17/12]$.} Interestingly, our result~\eqref{H21final} can be compared directly with accurate numerical relativity calculations, such as those in~\cite{varma2014gravitational, bustillo2015comparison}.

\section{Test of the current quadrupole with a constant shift}
\label{sec:shift}

In this section, we show that our final expression for the 3PN current quadrupole moment passes (one aspect of) the test of constant shifts. By this, we mean that when the two trajectories of the particles are shifted by $y_1^i\longrightarrow y_1^i + \epsilon^i$ and $y_2^i\longrightarrow y_2^i + \epsilon^i$, where $\epsilon^i$ denotes an infinitesimal constant purely spatial vector, the variation of the moment obeys the expected law of transformation of the moment under the shift to first order in that shift. 

We consider the case of the moments of a general isolated system made of an extended smooth matter distribution. There is then no need of UV regularisation and we may focus on the moments $I_L$ and $J_L$ in 3 dimensions, given by Eqs.~\eqref{moments3d}. The laws of transformation of such moments to first order in the shift have been found in \textit{linearized} gravity by Damour and Iyer~\cite{DI91a, DI91b} (see also~\cite{DI94}). In this case, the pseudo-tensor $\tau^{\mu\nu}$ reduces to the matter stress-energy tensor $T^{\mu\nu}$ with compact support, so that there is no necessity to resort to the IR regularisation with regulator $\tilde{r}^B$ and finite part when $B=0$. In this situation, the transformation laws read (see Appendix~B in~\cite{DI91a})\footnote{Here, the shift vector is denoted $\epsilon_i=\epsilon^i$, which contrasts with the usual Levi-Civita symbol $\varepsilon_{iab}=\varepsilon^{iab}$.}
\begin{subequations}\label{deltaIJlinear}
\begin{align}
\delta_\epsilon^\text{lin} I_L &= \ell \,\epsilon_{\langle i_\ell} I_{L-1\rangle} - \frac{4\ell}{c^2(\ell+1)^2} \,\epsilon_a J^{(1)}_{b\langle L-1} \,\varepsilon_{i_\ell\rangle ab} + \frac{(\ell-1)(\ell+3)}{c^2(\ell+1)^2(2\ell+3)} \,\epsilon_a \,I^{(2)}_{a L}\,,\label{deltaIlinear}\\
\delta_\epsilon^\text{lin} J_L &= \frac{(\ell-1)(\ell+1)}{\ell} \,\epsilon_{\langle i_\ell} J_{L-1\rangle} + \frac{1}{\ell} \,\epsilon_a I^{(1)}_{b\langle L-1} \,\varepsilon_{i_\ell\rangle ab} + \frac{(\ell-1)(\ell+3)}{c^2 \ell(\ell+2)(2\ell+3)} \,\epsilon_a \,J^{(2)}_{a L}\,.\label{deltaJlinear}
\end{align}\end{subequations}
Such laws follow from the irreducible decomposition of the metric and from the fact that the components of the matter tensor behave under the spatial constant shifts like scalars, \textit{i.e.}, ${T'}^{\mu\nu}(x') = T^{\mu\nu}(x)$, considering now the action of the shifts as a passive coordinate transformation ${x'}^\mu = x^\mu + \epsilon^\mu$ with $\epsilon^\mu=(0,\epsilon^i)$.

For the full non-linear theory, driven by Eqs.~\eqref{EE}--\eqref{tau}, one must take into account the non-linear gravitational source term $\Lambda^{\mu\nu}$. Now, even though the pseudo-tensor $\tau^{\mu\nu}$ behaves in the same way as the matter tensor under constant spatial shifts, \textit{i.e.}, ${\tau'}^{\mu\nu}(x') = \tau^{\mu\nu}(x)$, the transformation laws~\eqref{deltaIJlinear} obeyed by the moments $I_L$ and $J_L$ are then expected to be modified, notably because those involve the regularisation factor $\tilde{r}^B$ dealing with the fact that the pseudo-tensor is no longer with compact support. This \emph{a priori} implies that the linear transformation laws~\eqref{deltaIJlinear} must be augmented by certain non-linear corrections, which may be referred to as $\delta_\epsilon^\text{nonlin} I_L$ and $\delta_\epsilon^\text{nonlin} J_L$. 

Ignoring all odd powers of $1/c$, we have found that the mass quadrupole moment $I_{ij}$ at 3PN order does satisfy the linear transformation law~\eqref{deltaIlinear}, which means that the non-linear correction in this case actually happens to start only at the 4PN order:
\begin{align}\label{deltanonlinIij}
\delta_\epsilon^\text{nonlin}I_{ij} = \mathcal{O}\left(\frac{1}{c^8}\right)\,.
\end{align}
However, in the case of the current quadrupole moment $J_{ij}$ (neglecting again possible odd-type contributions at 2.5PN order), we find that the non-linear contribution arises precisely at the 3PN order. Since it is due to the fact that the pseudo tensor has a non-compact support, we expect this non-linear correction to be made of some combination of the multipole moments parametrizing the expansion of the metric at infinity. Looking at the only possible contribution at that order, we infer from dimensional analysis that the non-linear correction is necessarily of the type
\begin{align}\label{deltanonlinJij}
\delta_\epsilon^\text{nonlin}J_{ij} = \eta\,\frac{G^2 M^2}{c^6} \,\epsilon^{a}\,\varepsilon_{ab\langle i}\,I^{(3)}_{j\rangle b} + \mathcal{O}\left(\frac{1}{c^8}\right) \,,
\end{align}
where $I_{ij}$ is the (Newtonian here) quadrupole moment and the other factors comprise two masses $M$ so that the interaction is cubic. We have introduced an unknown numerical coefficient $\eta$ in front. Now, as an important though partial check of our final result for the 3PN current-type quadrupole $J_{ij}$, we have verified that it satisfies the law of transformation under the shift if and only if the numerical coefficient in the non-linear term~\eqref{deltanonlinJij} is $\eta=\frac{58}{105}$.

Although we have not determined the coefficient $\eta$ from scratch, the latter verification is enough for our purpose. Indeed, it is straightforward to see that any offending term in the quadrupole moment itself that is not checked by the fact that we have not determined the coefficient $\eta$ is necessarily of the type
\begin{align}\label{deltanonlinJij2}
\Delta J_{ij} = \eta'  \,\frac{G^2M^2}{c^6}J^{(2)}_{ij}\quad\Longrightarrow\quad\delta_\epsilon \Delta J_{ij} = \frac{\eta'}{2} \frac{G^2 M^2}{c^6} \,\epsilon^{a}\,\varepsilon_{ab\langle i}\,I^{(3)}_{j\rangle b}\,.
\end{align}
Now, such term $\Delta J_{ij}$ does not vanish in the test-mass limit $\nu\to 0$ and, therefore, its coefficient $\eta'$ has already been verified by the correct perturbative limit, which we have checked independently from black-hole perturbation theory~\cite{TSasa94}. We thus conclude that the above partial test of the constant shifts together with the perturbative limit grant us with a satisfying level of confidence in our result. 

\acknowledgements
It is a pleasure to thank Fran\c{c}ois Larrouturou for very interesting regular discussions.

\appendix

\section{Compendium of formulas for the irreducible decomposition} 
\label{sec:Young}

\begin{itemize}
	\item
Decomposition of a tensor $\mathcal{T}^i_L$ which is STF in the indices $L$ in irreducible tensors
\begin{align}
\mathcal{T}^i_L  = \mathcal{T}\indices{^{(i}_{L)}}
+ \frac{2\ell}{(\ell+1)} \mathop{\mathcal{S}}_{L} \mathcal{T}\indices{^{[i}_{i_\ell]L-1}}\, .
\end{align}
\item
Decomposition of a tensor $\mathcal{T}\indices{^{ij}_L}$ which is STF in $L$ and in $ij$ in irreducible tensors
\begin{align}
\mathcal{T}\indices{^i^j_L}  = \mathcal{T}\indices{^{(i}^j_{L)}}
+ \frac{4(\ell+1)}{(\ell+2)}
\mathop{\mathcal{S}}_{ij,\, L} \mathop{\mathcal{A}}_{i i_1}
\mathcal{T}\indices{^i^{(j}_{L)}} + \frac{4(\ell-1)}{(\ell+1)} \mathop{\mathcal{S}}_{L}
\mathcal{T}\indices{^{[i}_{i_1]}^{[j}_{i_2}_L}\, .
\end{align}
\item 
Decomposition of a tensor $\mathcal{F}\indices{^i_L}$ which is STF in $L$
\begin{align}
\mathcal{F}\indices{^i_L} &= \hat{\mathcal{F}}\indices{^i_L}
+ \frac{\ell (2\ell+d-4)}{(\ell+d-3)(2\ell+d-2)} \delta_{i\langle i_\ell}
\mathcal{H}_{L-1\rangle} \nonumber \\ & = \hat{\mathcal{F}}\indices{^i_L}
+ \frac{\ell (2\ell+d-4)}{(\ell+d-3)(2\ell+d-2)} \delta_{i(i_\ell}
\mathcal{H}_{L-1)} - \frac{\ell (\ell-1)}{(\ell+d-3)(2\ell+d-2)}
\delta_{(i_\ell i_{\ell-1}} \mathcal{H}_{L-2)i}\, ,
\end{align}
where $\displaystyle \hat{\mathcal{F}}\indices{^i_L} = \mathop{\text{TF}}_{iL} \,\mathcal{F}\indices{^i_L}$ and $\mathcal{H}_{L-1} = \mathcal{F}\indices{^k_{kL-1}}$.
\item
Inverse formula
\begin{align}
\hat{\mathcal{F}}\indices{^i_L} &= \mathop{\text{TF}}_{iL} \mathcal{F}\indices{^i_L}
\equiv \mathcal{F}\indices{^i_L} - \frac{\ell (2\ell+d-4)}{(\ell+d-3)(2\ell+d-2)}
\delta_{i\langle i_\ell} \mathcal{F}\indices{^k_{L-1\rangle k}} \\
& = \mathcal{F}\indices{^i_L} - \frac{\ell (2\ell+d-4)}{(\ell+d-3)(2\ell+d-2)}
\delta_{i(i_\ell} 
\mathcal{F}\indices{^k_{L-1)k}} + \frac{\ell (\ell-1)}{(\ell+d-3)(2\ell+d-2)}
\delta_{(i_\ell i_{\ell-1}} \mathcal{F}\indices{^k_{L-2)ik}} \,.\nonumber
\end{align}
\item
Decomposition of a tensor $\mathcal{F}\indices{^i^j_L}$ which is STF in $L$ and $ij$ into TF tensors
\begin{align}
\mathcal{F}\indices{^i^j_L} &= \hat{\mathcal{F}}\indices{^{ij}_L}
+ \frac{2\ell (2\ell+d -4)}{(\ell+d-2)(2\ell+d-2)}
\mathop{\text{STF}}_{ij,\, L} \delta^i_{i_\ell} \bigg[ \mathcal{H}\indices{^j_{L-1}} +
\frac{4\ell}{(d-2)(2\ell+d)} \mathcal{H}\indices{^{(j}_{L-1)}} \bigg]
\nonumber \\ & \quad + \frac{\ell (\ell-1) (2\ell+d-6)}{(\ell+d-4)(\ell+d-3)(2\ell+d-2)}
\mathop{\text{STF}}_{ij,\, L} \delta^i_{(i_\ell} \delta^j_{i_{\ell-1}}
\mathcal{L}_{L-2)}  \nonumber \\
&= \hat{\mathcal{F}}\indices{^{ij}_L}
+ \frac{2\ell (2\ell+d -4)}{(\ell+d-2)(2\ell+d-2)}
\mathop{\mathcal{S}}_{ij,\, L} \delta^i_{i_\ell} \bigg[ \mathcal{H}\indices{^j_{L-1}} +
\frac{4\ell}{(d-2)(2\ell+d)} \mathcal{H}\indices{^{(j}_{L-1)}} \bigg]
\nonumber \\ & \quad  - \frac{2\ell (\ell -1) }{
	(\ell+d-2)(2\ell+d-2)}  \mathop{\mathcal{S}}_{ij,\, L} \delta_{i_{\ell}
	i_{\ell-1}} \bigg[ \mathcal{H}\indices{^i_{L-2j}} + \frac{4
	\ell}{(d-2)(2\ell+d)} \mathcal{H}\indices{^{(i_{\ell-2}}_{L-2j)}} \bigg]
\nonumber \\ &\quad - \frac{2\ell (2\ell+d-4)}{(d-2)(\ell+d-2)(2\ell+d)}
\delta^{ij} \mathcal{H} \indices{^{(i_\ell}_{L-1)}} \nonumber \\
&\quad + \frac{\ell (\ell-1) (2\ell+d-6)}{(\ell+d-4)(\ell+d-3)(2\ell+d-2)}
\delta^i_{(i_\ell} \delta^j_{i_{\ell-1}} \mathcal{L}_{L-2)}  \nonumber \\
&\quad - \frac{2\ell
	(\ell-1)(\ell-2)(2\ell+d-6)}{(\ell+d-4)(\ell+d-3)(2\ell+d-4)(2\ell+d-2)}
\mathop{\mathcal{S}}_{ij} \delta^i_{(i_\ell} \delta_{i_{\ell-1} i_{\ell-2}}
\mathcal{L}_{L-3)j} \nonumber \\
&\quad + \frac{\ell(\ell-1)(\ell-2)(\ell-3)}{(\ell+d-4)(\ell+d-3)(2\ell+d-4)(2\ell+d-2)}
\delta_{(i_\ell i_{\ell-1}} \delta_{i_{\ell-2} i_{\ell-3}}
\mathcal{L}_{L-4)ij} \nonumber \\
& \quad- \frac{\ell(\ell-1)(2\ell+d-6)}{(\ell+d-4)(\ell+d-3)(2\ell+d-4)(2\ell+d-2)}
\delta^{ij} \delta_{(i_{\ell}i_{\ell-1}} \mathcal{L}_{L-2)} \, ,
\end{align}
where $\displaystyle \hat{\mathcal{F}}\indices{^i^j_L}= \mathop{\text{TF}}_{ijL} \mathcal{F}\indices{^{ij}_{L}}$, $\displaystyle \mathcal{H}\indices{^i_{L-1}} =
\mathop{\text{TF}}_{iL-1}\mathcal{F}\indices{^{ik}_{kL-1}}$ and $\mathcal{L}_{L-2} = \mathcal{F}\indices{^{kl}_{klL-2}}$.
\item
Number ($\sharp$) of independent components of irreducible tensors
\begin{subequations}
\begin{align}
& \sharp(\mathcal{H}^{[i}_{\phantom{[i}i_\ell]L-1}) =
\sharp(\mathcal{K}^{[i}_{\phantom{[i}i_\ell]L-1})=  
\frac{(2\ell +d-2)(\ell+d-2) (d-2)_{\ell-1}}{(\ell+1)(\ell-1)!}\, ,\\
& \sharp(\mathcal{P}^{[i\phantom{i_{\ell-1}]}[j}_{\phantom{[i} i_{\ell-1}]  
	\phantom{[j} i_{\ell}] L-2}) = 
\frac{(2\ell+d-2)(\ell+d-1)(d-3)d (d-1)_{\ell-2}}{2\ell(\ell+1)(\ell-2)!} \, ,\label{eq:nbcompP} \\
& \sharp(\text{STF tensor of rank $\ell$}) = \frac{(2\ell+d-2) (d-1)_{\ell-1}}{\ell!}\, ,
\end{align}\end{subequations}
with the standard notation, \textit{e.g.}, $d_{\ell} \equiv d(d+1) \cdots (d+\ell-1)$, for the Pochhammer symbol.
\end{itemize}

\bibliography{ListeRef_Quadcourant.bib}

\end{document}